\documentclass[11pt]{article}
\usepackage {fullpage}						% set all margins to 1.5cm
\usepackage {color}

\usepackage {ucs}
\usepackage{upgreek} 						% to get upright \mu
\usepackage{mathbbol}
\usepackage [utf8x]{inputenc}
\usepackage {graphicx}

\usepackage {amsfonts,amsmath}%,amsthm,amssymb}

\usepackage {multicol}						% to locally have two columns
\usepackage {datetime} 						% date and time in latex
\usepackage {natbib}						% several bib styles & setting bibspace	\setlength{\bibsep}{0.3em}
\usepackage {subfigure}						% to have enumerated subfigures in one figure
\usepackage {parskip}						% no indentation and space between paragraphs
\usepackage {setspace}						% to set the vertical spacing
\usepackage[font=small,textfont=it,width=0.97\textwidth]{caption}		% to set the fonts for the captions, to use caption*
\usepackage {mathtools}						% to have dcases environment
\usepackage {array}							% to vertically align within a table and give width
\usepackage{tabularx}
\usepackage[normalem]{ulem}							% wide variety of underline
\usepackage{cancel}							% wide variety of underline

\renewcommand {\vec}{\mathbf}

\newcommand{\snodes}{s_\text{nodes}}
\newcommand{\anodes}{\alpha_\text{nodes}}
\renewcommand{\a}{\alpha}

	\newtheorem{experiment}{\textbf{Experiment}}[section]

\renewcommand{\(}{\left(}
\renewcommand{\)}{\right)}

\graphicspath{{./input/}}

\def\subtext#1{\text{#1}}

\newcommand{\change}[2]{#2}

\title{Stability, convergence, and sensitivity analysis of the FBLM and the corresponding FEM}
\author{
	N. Sfakianakis\thanks{
		Institute of Applied Mathematics, Heidelberg University, Im Neuenheimer Feld 205, 69120, Heidelberg, Germany %\email{sfakiana@math.uni-heidelberg.de}	
	}
	\and
	A. Brunk\thanks{
	Institute of Mathematics, Johannes Gutenberg-University, Staudingerweg 9, 55128, Mainz, Germany,  %\email{aabrunk@students.uni-mainz.de}
	}
	}
\begin{document}
\maketitle

\begin{abstract}
	This paper focuses on the study of the \textit{filament based lamellipodium model} (FBLM) and the corresponding \textit{finite element method }(FEM) from a numerical point of view. We study fundamental numerical properties of the FEM and justify the further use of the FBLM. \change{In particular:}{} {We exhibit that} the FEM satisfies a timestep stability condition that is consistent with the nature of the problem. We propose a particular strategy to automatically adapt the time step of the method. We show that the FEM convergences with respect to the (two-dimensional) space discretization in a series of characteristic and representative experiments. We embed and couple the FBLM with a complex extracellular environment comprised of chemical and haptic components and study their combined time evolution. {Under this prism}, we study the sensitivity of the FBLM on several of its controlling parameters and discuss their influence in the development of the model.
\end{abstract}

%-----------------------------------------------	---------------------------	
%--------------------------------------------------------------------------
\section{Introduction} 

\textit{Fibroblasts}, \textit{keratocytes}, \textit{cancer cells}, and other types \change{}{of} fast moving cells exhibit a particular crawling-like motion in which the \textit{lamellipodium} of the cells plays a pivotal role, \cite{Small2002,Svitkina1997,Yam2007,Postlethwaite1987,Gerisch1981,Iijima2002,Zigmond1973}. 

The lamellipodium is a sheet-like dense network \change{of linear \textit{biopolymers} of the protein \textit{actin} ---termed \textit{actin-filaments} or simply \textit{filaments}--- that can be found in the propagating front of the cells}{that can be found in the propagating front of the cells and is comprised of linear biopolymers of the protein \textit{actin}  ---termed \textit{actin-filaments} or simply \textit{filaments}.}. These {actin-filaments} are highly dynamic; they continuously \textit{polymerize}, \textit{adhere} to the substrate, and are subject to numerous other processes like  \textit{nucleation}, \textit{fragmentation}, \textit{capping}, and more \cite{Blanchoin_Plastino2014,Lauffenburger1996, Gittes1993, Tojkander2012,Mitchison1996,Jay1995,Chen1981}.

These processes affect the structure and \change{}{the} functionality of the lamellipodium and the motility of the cell, \cite{Small2002}. They are \change {greatly} influenced\change{}{, to a large extent, } by the extracellular environment, \change{e.g.}{} its chemical composition and the \textit{architecture} of the \textit{Extracellular Matrix} (ECM). The response of the cell to gradients of extracellular \textit{chemical signals} and ECM-bound \textit{adhesion sites} is termed \textit{chemotaxis} and \textit{haptotaxis} respectively.

There have been several approaches in the literature to model and simulate lamellipodium driven cell motility. In the current paper we do not aim to develop \change{and}{} a new model or to perform biologically relevant numerical experiments, hence we do not discuss the literature in detail\change{s}{} or compare the different model\change{}{s}. We merely refer the interested reader to some of \change{}{the} existing works \cite{Schwarz2010, Stevens2015, Voigt2015, Mogilner2009, Alt2009, DeSimone2011,Preziosi2013, Merkel2012,Madzvamuse2013,Ambrosi2016}. 

\change{In this work we follow}{The model we follow is} the \textit{Filament Based Lamellipodium Model} (FBLM), a two dimensional, two phase model that describes the lamellipodium at the level of actin-filaments. \change{This model}{It} was first introduced in \cite{Oelz2008,Oelz2010a} and was further extended in \cite{MOSS-model}. When endowed with a particular problem specific \textit{Finite Element Method} (FEM), the \change{}{the resulting } FBLM-FEM is able to reproduce a realistic, crawling-like lamellipodium driven motility \cite{MOSS-numeric,Brunk2016}.

Although the FBLM describes the dynamics of \change{}{the} actin-filaments and the lamellipodium, the deduced motility is understood as the motility of the full cell. This is \change{mostly}{} due to the fact that \change{in the case}{the role of the lamellipodium in the motility of the} model biological cell (\textit{fish keratocyte}) \change{that we consider, the role of the lamellipodium in the motility}{} is predominant, \cite{Small2002}. We henceforth will not distinguish between the two, and for convenience we will use the term \textit{cell motility} for both cases.

The FBLM and the corresponding FEM have been used so far in several works to simulate \change{some}{various} cases of cell motility, e.g. \cite{MOSS-model,MOSS-numeric,Brunk2016}. To date though no numerical \change{analysis}{} investigations have been presented in the literature to verify that the FBLM-FEM combination satisfies some (at least minimum) numerical prerequisites and justify thusly its further use in biological relevant situations. What moreover is missing from the relevant literature is a study of the corresponding \change{parameter set,}{parameters} and of the effect \change{that}{} they have \change{in}{on} the dynamics of the FBLM. Such parameter study would facilitate the \textit{parameter identification} procedure\change{(-s)}{} in the reproduction/simulation of realistic biological experiments. The current work aims to \change{\textit{partially fill}}{partially fill} these gaps and is split in two main parts\change{:}{.}

In the first part of the paper, we identify appropriate timestep \textit{stability} conditions in chemotaxis and haptotaxis experiments where we propose a relation between the time and the (two-dimensional) \change{space}{``space''} discretization steps. We investigate the dependence of the stability constant on the gradient of the ECM and propose an \change{\textit{Adaptive Time Control}}{\textit{adaptive time control}} (ATC) method for the \change{automated identification}{automatic adaptation} of the timestep of the \change{}{numerical} method. 

We also exhibit the \textit{convergence} of the FEM. For that we consider three generic and representative experiments and a \change{}{discretization} grid that is refined with respect to both \change{the}{} ``spatial'' variables. \change{The benefit of this method stems} 

As the model and the \change{}{numerical} method are quite complex, they do not allow for a rigorous numerical analysis study\change{. As}{; as} a result we restrict our work here to the experimental and case dependent study of these properties\change{,}{} and postpone the more rigorous numerical and analytical investigations for a separate work. Nevertheless, the benefit that stems from the first part of our work is \change{that we have verified}{mainly the verification} that the FBLM-FEM can be further employed in biologically relevant numerical simulations. Most notably, that the refinement of the mesh and the adaptation of the timestep will \change{}{be able to} reveal the dynamics of the model.

In the second part of the paper, we make one more step towards biological realism and embed the FBLM in a complex and adaptive chemical and {haptic} extracellular environment. Such extended FBLM-environment \change{models}{model combination} will allow us to better reproduce \textit{in vitro} biological experiments. The model for the environment we propose here is minimal. It includes only some basic components \change{and}{of the extracellular environment and we} use it primarily to present the \change{}{FBLM-environment} coupling\change{ between the FBLM and the environment}{}. 

Based on the \change{combination of the FBLM with the environment,}{FBLM-environment combination,} we perform a sensitivity analysis \change{of the FBLM}{} in a generic experimental case, where we identify the most influential parameters and address the effect they have in the dynamics of the model. We get this way a critical insight \change{in}{of} the different components and dynamics of the FBLM and pave the way for \change{}{more detailed and problem specific} parameter identification \change{works}{investigations} and the simulation of biologically relevant experimental scenarios. 

The structure of the paper is as follows: in Section \ref{sec:model} we describe briefly the FBLM and give some details on extensions of the model that we consider for the first time. In Sections \ref{sec:stbility} and \ref{sec:converge} we study the \textit{stability} and the \textit{convergence} of the the FEM. %In particular, we identify the maximal timestep of the method as a function of the space steps and other parameters of the problem, and identify the dependence of the stability constant on the gradient of the ECM. As is typically done in complex models and methods when no rigorous numerical analysis is conducted, both the stability and the convergence study of the method is exhibited in characteristic and representative experiments.
In Section \ref{sec:env} we introduce the model for the extracellular environment, and in Section \ref{sec:sensit}, we study the sensitivity of the FBLM-FEM on a series of its controlling parameters. %We identify the most influential ones, get an insight on the dynamics of the FBLM, and prepare this way the parameter identification procedures needed for reproducing realistic experimental scenarios. \remm{again}
In the Appendix we provide some basic information on the FEM and the FV methods we use to solve the FBLM and the environment.
%--------------------------------------------------------------------------	
%--------------------------------------------------------------------------
\section{The FBLM} \label{sec:model}
We present here some information on the FBLM and the new components that we include, and refer to \cite{Oelz2008,Oelz2010a,Schmeiser2010, MOSS-model,MOSS-numeric, Brunk2016} for more details.

The FBLM is a two-dimensional, two-phase continuum model that describes the dynamics of the lamellipodium by retaining key biological processes of the actin-filaments\change{ and the their}{, the} interactions\change{. The model distinguishes between two families (phases) of filaments and includes the interactions between them}{with each other}, as well as \change{between the network of the filaments and}{their interactions with} the extracellular environment. \change{In more details:}{}

The main assumptions behind the FBLM are \change{that}{a)} the lamellipodium is a two dimensional structure, and \change{that}{b)} the actin filaments are organized in two locally parallel families (denoted here by the superscripts $+$ or $-$). Each family covers a region with the filaments connecting the membrane with the inside of the cell. 

The filaments of the $\pm$ family are indexed by $\alpha\in [0,2\pi)$ and have a maximal length $L^\pm(\alpha,t)$ at time $t$. The two families are parametrized with respect to their arclength as
\begin{equation} \label{eq:arclength}
	\left\{\vec{F}^\pm(\alpha,s,t): -L^\pm(\alpha,t)\le s \le 0\right\}\subset {\mathbb R}^2,
\end{equation}
and coincide at their outer boundaries ($s=0$) with the membrane of the cell
\begin{equation}\label{eq:tether}
	\left\{\vec{F}^+(\alpha,0,t): 0\le \alpha < 2\pi\right\} = \left\{\vec{F}^-(\alpha,0,t): 0\le \alpha < 2\pi\right\},  \quad \forall\, t\geq 0 \,.
\end{equation}
They moreover satisfy the constraint 
\begin{equation}\label{eq:inext}
\left|\partial_s \vec{F}^\pm(\alpha,s,t)\right| = 1 \quad\forall\, (\alpha,s,t) \;,
\end{equation}
that is understood as an \textit{inextensibility} condition between their monomers.

We assume that filaments of the same family do not intersect each other
\begin{equation}
\det \(\partial_\a \vec F^\pm, \partial_s \vec F^\pm\)>0
\end{equation}
and that filaments of different families cross at most once
%\begin{equation}
%	\left\{ s^\pm = s^\pm(\alpha^+,\alpha^-,t)\Big |~ \exists \a^+,\a^-\, :\, \vec{F}^+(\alpha^+,s^+,t) = \vec{F}^-(\alpha^-,s^-,t)\right\}.
%\end{equation}
%
\begin{equation}
	\Big\{ \forall (\a^+,\a^-)\ \exists \text{ at most one } (s^+,s^-) \text{ such that }\, \vec{F}^+(\alpha^+,s^+,t) = \vec{F}^-(\alpha^-,s^-,t)\Big\}.
\end{equation}

%As a consequence of the above assumptions, there exist coordinate transformations 
%$\psi^\pm:\, (\alpha^\mp,s^\mp) \mapsto (\alpha^\pm,s^\pm)$ such that 
%\[
%\vec{F}^\mp = \vec{F}^\pm \circ \psi^\pm \;. \comm{\text{do we need it?}}
%\]

\begin{figure}[t]
	\begin{center}
		%		{\includegraphics[width=0.65\linewidth]{input/DietmarImage1.pdf}}\\
		\begin{picture}(220,88)
		\put(0,0){\includegraphics[width=22em]{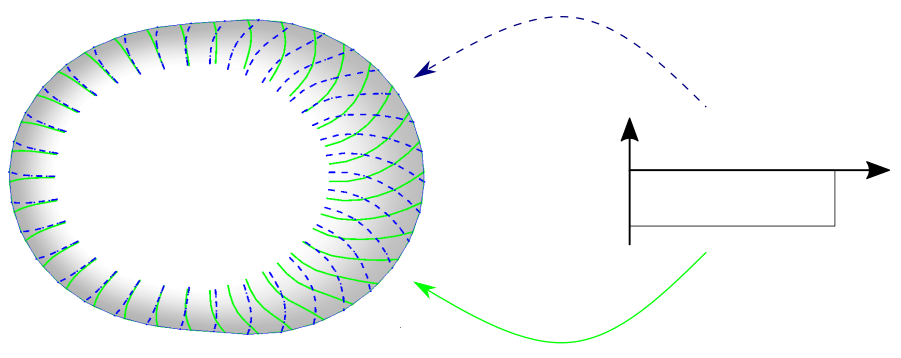}}
		%		\put(110,0){\scriptsize.}
		%		\put(110,85){\scriptsize.}
		%		\put(61,42){\scriptsize$\Omega(t)$}
		\put(117,65){\scriptsize$F^+$}
		\put(117,12){\scriptsize$F^-$}
		\put(175,37){\scriptsize$B_0$}
		\put(147,41){\scriptsize$0$}
		\put(141,28){\scriptsize$-1$}
		\put(198,48){\scriptsize$2\pi$}
		\put(212,37){\scriptsize$\a$}
		\put(157,53){\scriptsize$s$}
		\end{picture}
	\end{center}
	\caption{Graphical representation of the $\vec F^\pm$ that map the domain of dependence $B_0$ \eqref{eq:rescaled.B} to the lamellipodium. The $s=0$ boundary of $B_0$ is mapped to the membrane of the cell and the $s=-1$ to the minus-ends of the filaments inside the cell. The filaments and the other functions of $\a$ are periodic with respect to $\a$. The ``filaments'' plotted in the lamellipodium correspond to the discretization interfaces of $B_0$ along the $\a$ direction.}\label{fig:domains}
\end{figure}
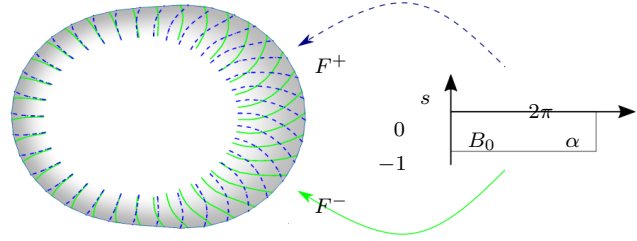

In the heart of the FBLM is found the system of equations
\begin{align} 
0=\underbrace{\mu^B \partial_s^2\left(\eta \partial_s^2 \vec{F}\right)}_\subtext{bending} &-\underbrace{\partial_s\left(\eta \lambda_{\rm inext} \partial_s \vec{F}\right)}_\subtext{in-extensibility} +\underbrace{\mu^A \eta D_t \vec{F}}_\subtext{adhesion} \nonumber + \underbrace{\partial_s\left( p(\rho) \partial_\a \vec{F}^{\perp}\right)
	-\partial_\a \left( p(\rho) \partial_s \vec{F}^{\perp}\right)}_\subtext{pressure} \nonumber\\
&\pm\underbrace{\partial_s\left(\eta\eta^* \widehat{\mu^T} (\phi-\phi_0)\partial_s \vec{F}^{\perp}\right)}_\subtext{twisting}  
+ \underbrace{\eta\eta^* \widehat{\mu^S}\left(D_t \vec{F} - D_t^* \vec{F}^*\right)}_\subtext{stretching} \;,\label{eq:strong}
\end{align}
where $\vec{F}^\bot = (F_1,F_2)^\bot = (-F_2,F_1)$ and where we have dropped the $\pm$ notation and focus on one of the two families/equations. The other family is indicated by the superscript $^*$ for which a similar equation holds.

The function  $\eta(\alpha,s,t)$ represents the (number) density of filaments of length at least $-s$ at time $t$ with respect to $\alpha$. The corresponding submodels used to derive the evolution of $\eta$ (and $L(\alpha,t)$) incorporate the effects of polymerization, depolymerization, branching, and capping, see \cite{MOSS-model}. %We only note that faster polymerization (even locally) leads to wider lamellipodia. \comm{something on the haptotaxis as well?}

The first term on the right hand side of (\ref{eq:strong}) describes the resistance of the filaments against bending, the second term is a tangential tension force, stemming from the inextensibility constraint (\ref{eq:inext}) with the \textit{Lagrange multiplier} $\lambda_{\rm inext}(\alpha,s,t)$. The third term describes the friction between the filament network and the substrate.

The filaments polymerize at the leading edge with rate $v(\alpha,t)\ge 0$. The \textit{material derivative} operator
\[
	D_t  := \partial_t  - v\partial_s 
\]
describes the velocity of the actin-material relative to the substrate. In a similar way we set $D_t^\ast  := \partial_t  - v^\ast \partial_s$.

The pressure effect in (\ref{eq:strong}) is caused by Coulomb repulsion between neighbouring filaments of the same family with \textit{pressure} $p(\rho)$ given by the density of actin as
\begin{equation} \label{eq:rho}
	\rho = \frac{\eta}{\left|\det(\partial_\a \vec{F}, \partial_s \vec{F})\right|}  \;.
\end{equation}

The last two terms in (\ref{eq:strong}) model the interaction between the two families caused by \textit{elastic cross-links} and/or \textit{branch junctions}. The first  one describes the resistance against changing the angle 
\[
	\phi=\arccos (\partial_s \vec{F}\cdot\partial_s \vec{F}^*)
\]
between crossing filaments away from the equilibrium angle $\phi_0$ of the cross-linking molecule. The second one describes the friction between the two families analogously to the friction with the substrate. %The friction coefficients have the form
%\[
%	\widehat{\mu^{T,S}} = \mu^{T,S} \left| \frac{\partial\a^*}{\partial s} \right| \;.
%\]

%with $\mu^{T,S}>0$, and the partial derivative refers to the coordinate transformation
%$\psi^*$, which is also used when evaluating partial derivatives of $\vec{F}^*$.

The system (\ref{eq:strong}) is subject to the boundary conditions 
\begin{align} \label{eq:newBC}
	- \mu^B\partial_s\left(\eta\partial_s^2 \vec{F}\right) -&~ p(\rho)\partial_\a \vec{F}^\perp + \eta \lambda_{\rm inext} \partial_s \vec{F}  
		\mp\eta\eta^* \widehat{\mu^T}(\phi-\phi_0)\partial_s \vec{F}^\perp \\
		=&~\left\{
			\begin{array}{l l}
				\eta \left(f_{\rm tan}(\alpha)\partial_s \vec{F} + f_{\rm inn}(\alpha) \vec{V}(\alpha)\right), & \quad \mbox{for 	} s=-L \;,\\
				\pm\lambda_{\rm tether} \nu, & \quad  \mbox{for }  s=0 \;,
			\end{array}\right.\nonumber\\
		\eta \partial_s^2 \vec{F} =&~ 0, \qquad \mbox{for } s=-L,0 \;.\nonumber
\end{align}
The terms in the second line, describe forces applied to the filament ends. The force in the  direction $\nu$ orthogonal to the leading edge at $s=0$ arises from the constraint (\ref{eq:tether}) with the Lagrange parameter $\lambda_{\rm tether}$. The forces at the inner boundary $s=-L$ model the contraction effect of actin-myosin  interaction in the interior region, refer to \cite{MOSS-model} for details.

Fundamental to the motility of the cell, is the breaking of the symmetry in the thickness of the lamellipodium. This way, the effective pulling force becomes stronger in the direction of the wider lamellipodium \cite{Yam2007}. The maximal length of the filaments and width of the lamellipodium $L(\a,t)$,  depends on the polymerization rate. Based on the capping, severing, and filament nucleation procedures, we have deduced in \cite{MOSS-model} the relation
\begin{equation}\label{eq:width_and_v}
	L(\a,t)= \frac{\kappa_\text{cap}}{\kappa_\text{sev}} + \sqrt{\frac{\kappa_\text{cap}^2}{\kappa_\text{sev}^2} + \frac{2v(\a,t)}{\kappa_\text{sev}}\log{\frac{\eta(0,t)}{\eta_\text{min}}}},
\end{equation}
which reveals the direct dependence of the width of the lamellipodium $L(\a,t)$ to the polymerization rate of $v(\a,t)$ of the local filaments.

For more details on the FBLM we refer to \cite{Oelz2008,Oelz2010a,Schmeiser2010, MOSS-model,MOSS-numeric}.

\subsection*{Adjusting the polymerization rate}
We consider in this work two mechanisms to control the polymerization rate $v$ of the filaments. The first is the direct response of the intracellular polymerization mechanism to extracellular chemical signals as they are perceived by the cell through transmembrane receptors. The second mechanism represents (unspecified in this paper) intracellular processes that destabilize, cut off, or even enhance the response of the polymerization mechanism.

For the first mechanism we assume that the polymerization rate $v^\pm_\text{ext}(\a,t)$ of the filament $\a$ is adjusted between the minimum and the maximum value $v_{\min}$, $v_{\max}$ of the cell polymerization mechanism according to the density of the extracellular chemical signal $c$ (that serves as a chemo-attractant) by the formula
\begin{equation}\label{eq:chem.pol}
	v^\pm_\text{ext} (\a,t)= v_{\max} - (v_{\max} -v_{\min}) e^{-\lambda_\text{res} c^\pm (\a,t)},
\end{equation}
where $c^\pm (\a,t)$ is the density of the extracellular chemical $c$ at the barbed end of the filament $\a$ at time $t$, i.e
$$c^\pm (\a,t) = c\( \vec F^\pm (\a,0,t),t\).$$
The coefficient $\lambda_\text{res}$ represents the response of the cell and in particular of the polymerization mechanism to changes of the extracellular chemical. Larger $\lambda_\text{res}$ values lead to more pronounced changes of the polymerization rate and to more polarized cells. 

The exponential function in \eqref{eq:chem.pol} has no biological justification; it is used merely to provide a smooth transition from the minimum $v_{\min}$ to the maximum $v_{\max}$ polymerization rate in a continuous and controlled manner. Other functions with the same attributes could be used in its place. 

The second mechanism that we consider describes primarily intracellular processes. For biological reasons that are not specified in this work, the polymerization mechanism can be hampered or otherwise destabilized, leading to an assortment of phenomena like persistent very high or low polymerization rates, abrupt changes of the polymerization rate, etc. This part of the model was previously proposed in \cite{MOSS-model} where it was used to prescribe the polymerization rate directly on the membrane of the cell.

The conditions that destabilize the polymerization mechanism are important in a assortment of phenomena (pathological or not) which are beyond the scope of this paper, so we will not comment on them any more. We understand though the biological significance of both mechanisms as well as their distinctive functionality and use both \change{}{of} them in this work. 

Overall, the polymerization rate $v^\pm$ that we consider is given as 
\begin{equation}\label{eq:int.pol}
	v^\pm(\a,t) = \mathcal D_\text{stb}\(v^\pm_\text{ext}(\a,t)\),
\end{equation}
where $\mathcal D_\text{stb}$ describes the internal controlling mechanism that can potentially depend on a large number of cellular processes. Nevertheless, unless otherwise stated, we assume throughout this paper that $\mathcal D_\text{stb}=\text{id}$ and hence
\begin{equation}\label{eq:int.pol.II}
	v^\pm(\a,t) = v^\pm_\text{ext}(\a,t).
\end{equation}

For the numerical solution of the FBLM we employ a problem specific FEM that we briefly describe in Appendix \ref{sec:FEM}. It was previously developed in \cite{MOSS-numeric, Brunk2016}, where we refer for more details. 

%--------------------------------------------------------------------------	
%--------------------------------------------------------------------------
\begin{figure}[t]
	\centering
	\begin{tabular}{cc}
		\hskip -1em
		{\includegraphics[width=0.45\linewidth]{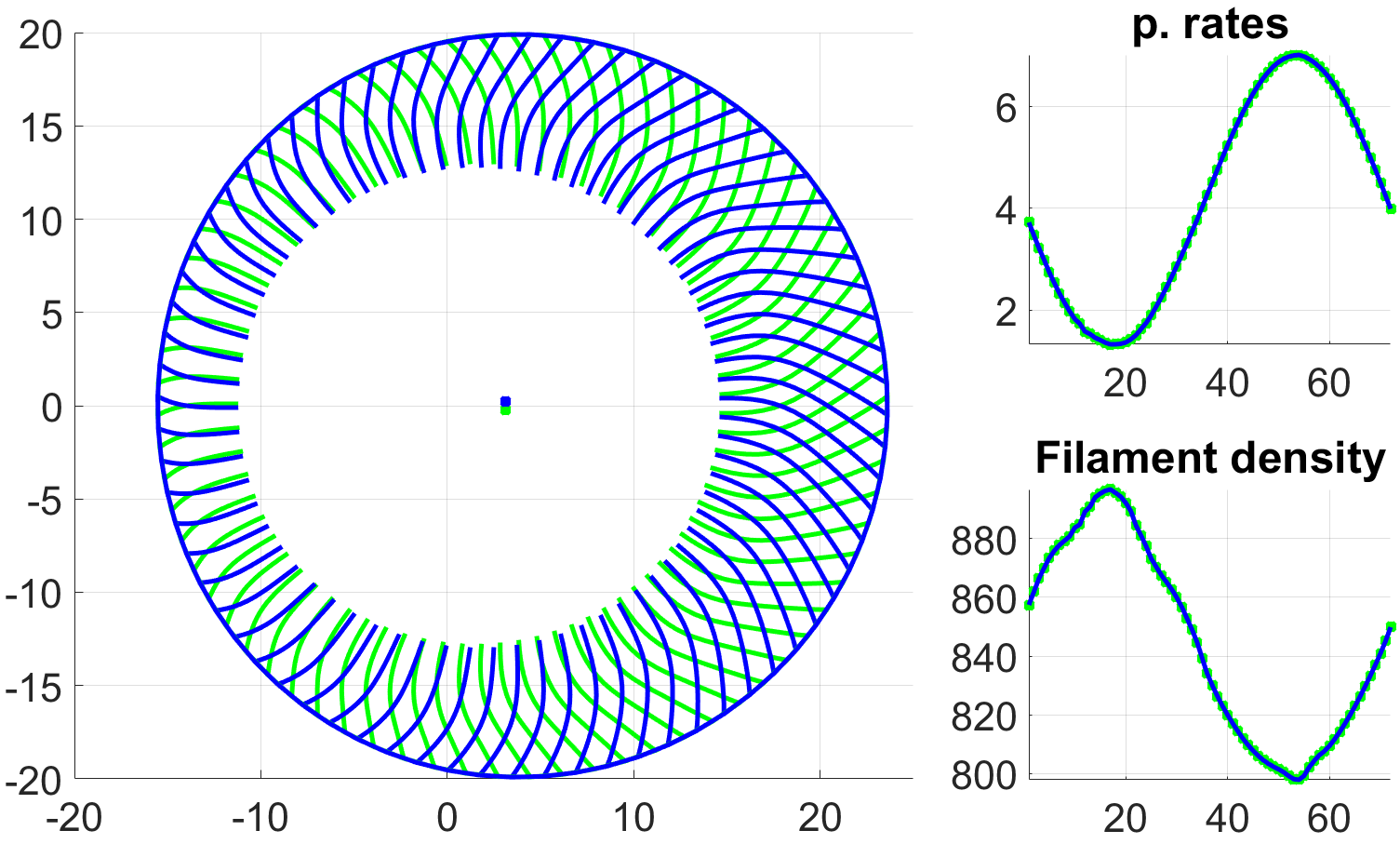}}
		&{\includegraphics[width=0.45\linewidth]{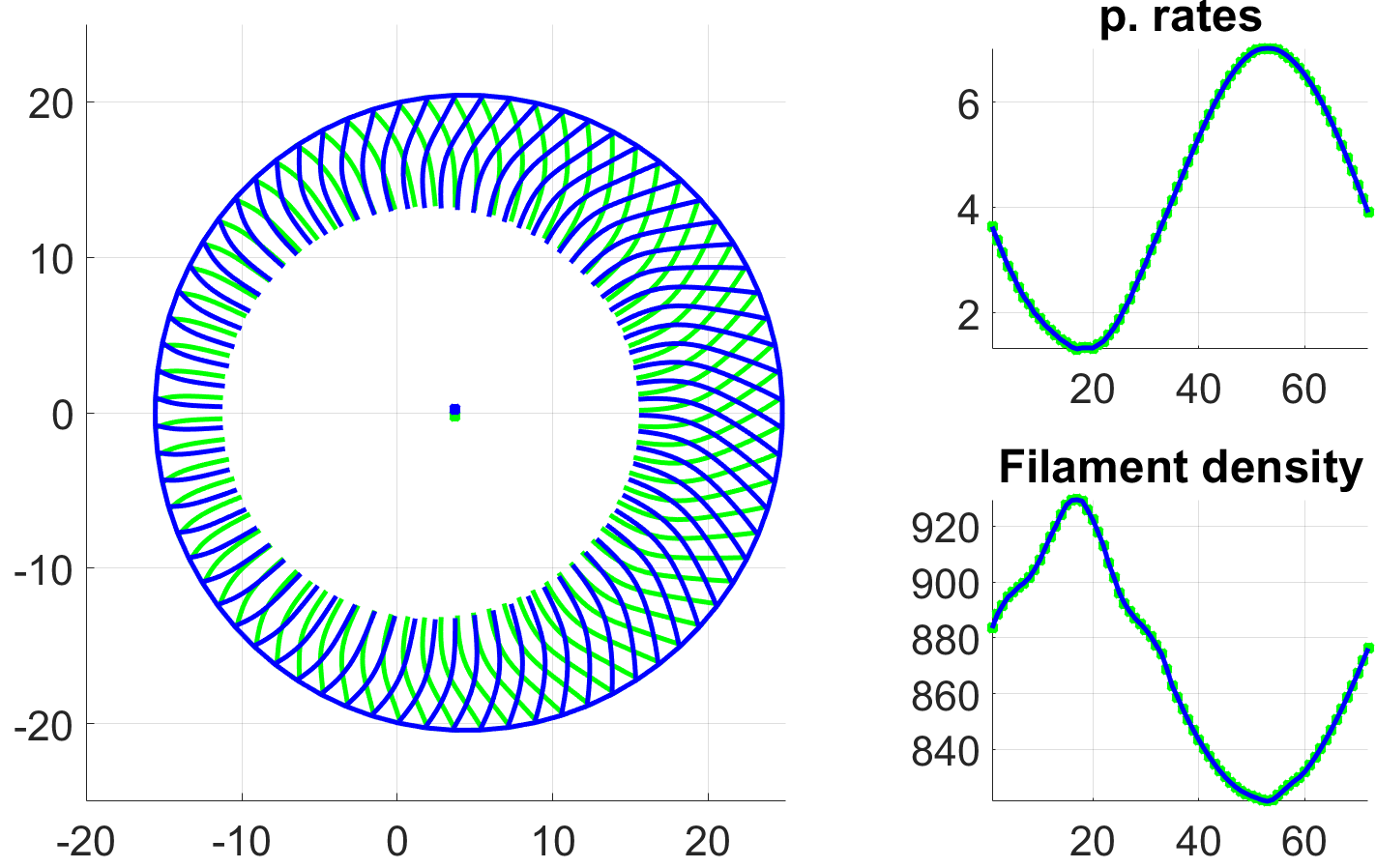}}\\
		\footnotesize{(a) $\Delta t = 0.0002$} & \footnotesize{(b) $\Delta t=0.001$}
	\end{tabular}
	\caption{Final time conformations for the Experiment \ref{exp:dt.chemo} with $\snodes=18$, $\anodes=72$. Showing in the large panel the full lamellipodium as comprised by the two families of ``discrete'' filaments (corresponding to the discretization lines of the domain), and in the smaller panels, the polymerization rates of the individual ``discrete'' filaments and the corresponding number density of the filaments at the leading edge with respect to their index. In both, the conformation of the lamellipodium and its discrete filaments, as well as the polymerization rates and filament densities are smooth.}\label{fig:18x72}
\end{figure}

\section{Timestep stability}\label{sec:stbility} 

The complexity of the FBLM \eqref{eq:strong} and the FEM \eqref{eq:FEM} do not allow for a rigorous stability analysis. Instead, we perform here an experimental/numerical investigation where we exhibit the existence of regions of stability for the timestep $\Delta t$ in terms of the space discretizaiton steps $\Delta s$ and $\Delta \a$.

To this end, we first note that the structure of the FEM \eqref{eq:FE:V}--\eqref{eq:I2D} indicates a particular relation between $\Delta t$, $\Delta s$, and $\Delta \a$, of the form:
\begin{equation}\label{eq:CFL}
	\Delta t \leq C\,(\Delta s)^3\,\Delta \a\,.
\end{equation}
To identify the stability constant $C$, we numerically solve indicative experiments for different combinations of $\Delta t$, $\Delta s$, $\Delta a$. For each combination, we characterize the resulting conformation as smooth (or not) and accordingly accept (or not) the corresponding combination. The largest $\Delta t$ to result to smooth solutions gives rise to $C$. As we see later, this procedure can be used to set the timestep of the method in an automated way.

The experiments that we consider are particular; one chemotaxis and one haptotaxis. In the fist we exhibit the approach we follow to identify the stability constraint, and discuss a computational approach to set the timestep automatically. In the second we go one step further and identify the relation between the stability constraint as the gradient of the ECM. 

The first experiment we consider is a chemotaxis driven cell migration.
\begin{experiment}[Stability -- Chemotaxis]\label{exp:dt.chemo}
	An initially rotational symmetric cell migrates under the influence of a chemical signal. The direction and strength of the signal and the final simulation time are chosen in a way that the deformation and migration of the cell is small, while at the same time the width of the lamellipodium (and hence the effective pulling force) becomes significantly asymmetric around the lamellipodium. %See e.g. Figures \ref{fig:18x72}--\ref{fig:5x36}.%, \ref{fig:9x72}, \ref{fig:5x36}.

	The parameters for this experiment are given in Table \ref{tbl:parameters}. The polymerization rate varies smoothly from the minimum value at the posterior side of the cell, to the maximum value at the anterior, see \cite{MOSS-model} for more details.
\end{experiment}

\begin{figure}[t]
	\centering
	\begin{tabular}{cc}
		\hskip -0.7em
		{\includegraphics[width=0.45\linewidth]{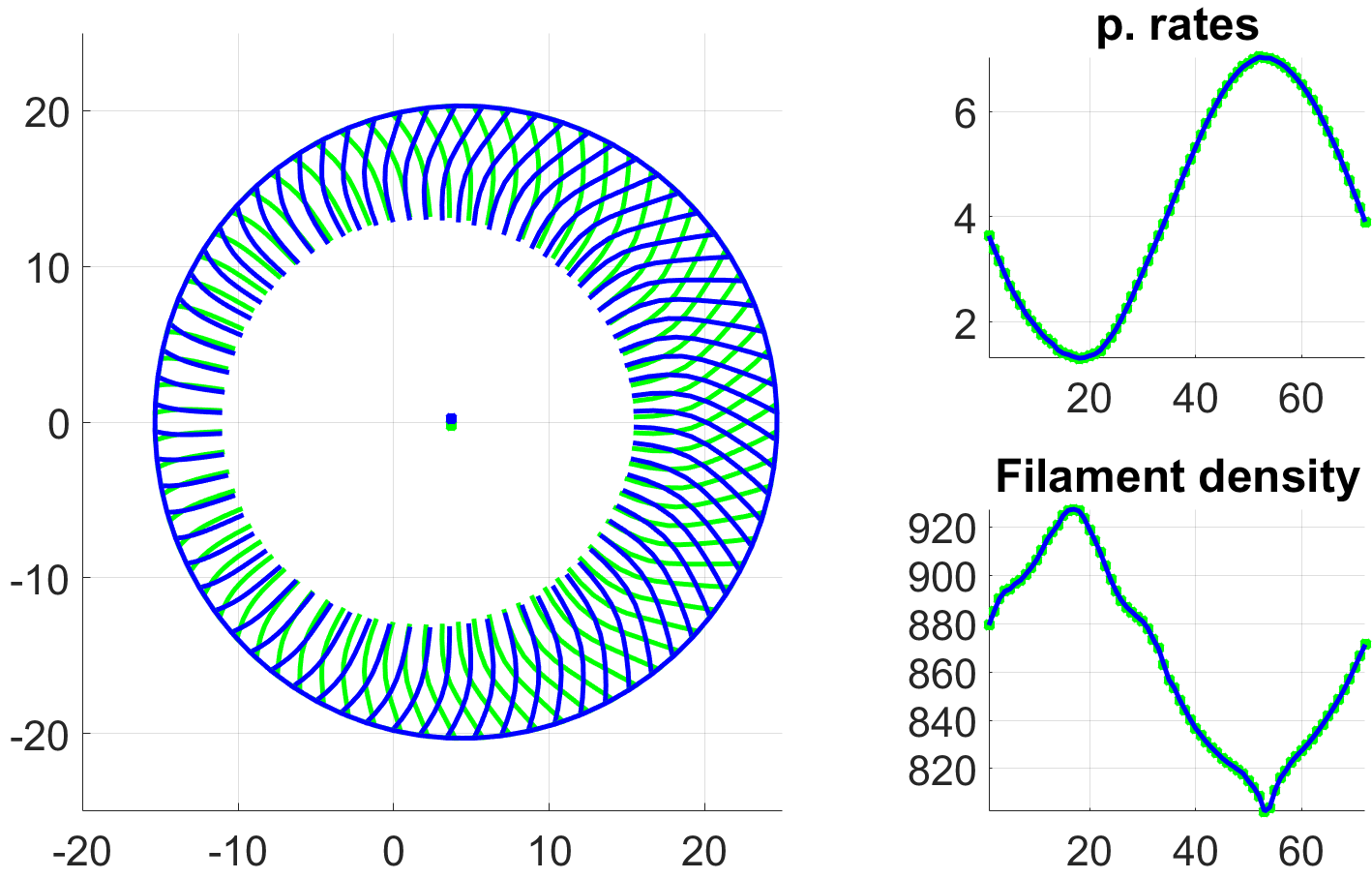}}
		&{\includegraphics[width=0.45\linewidth]{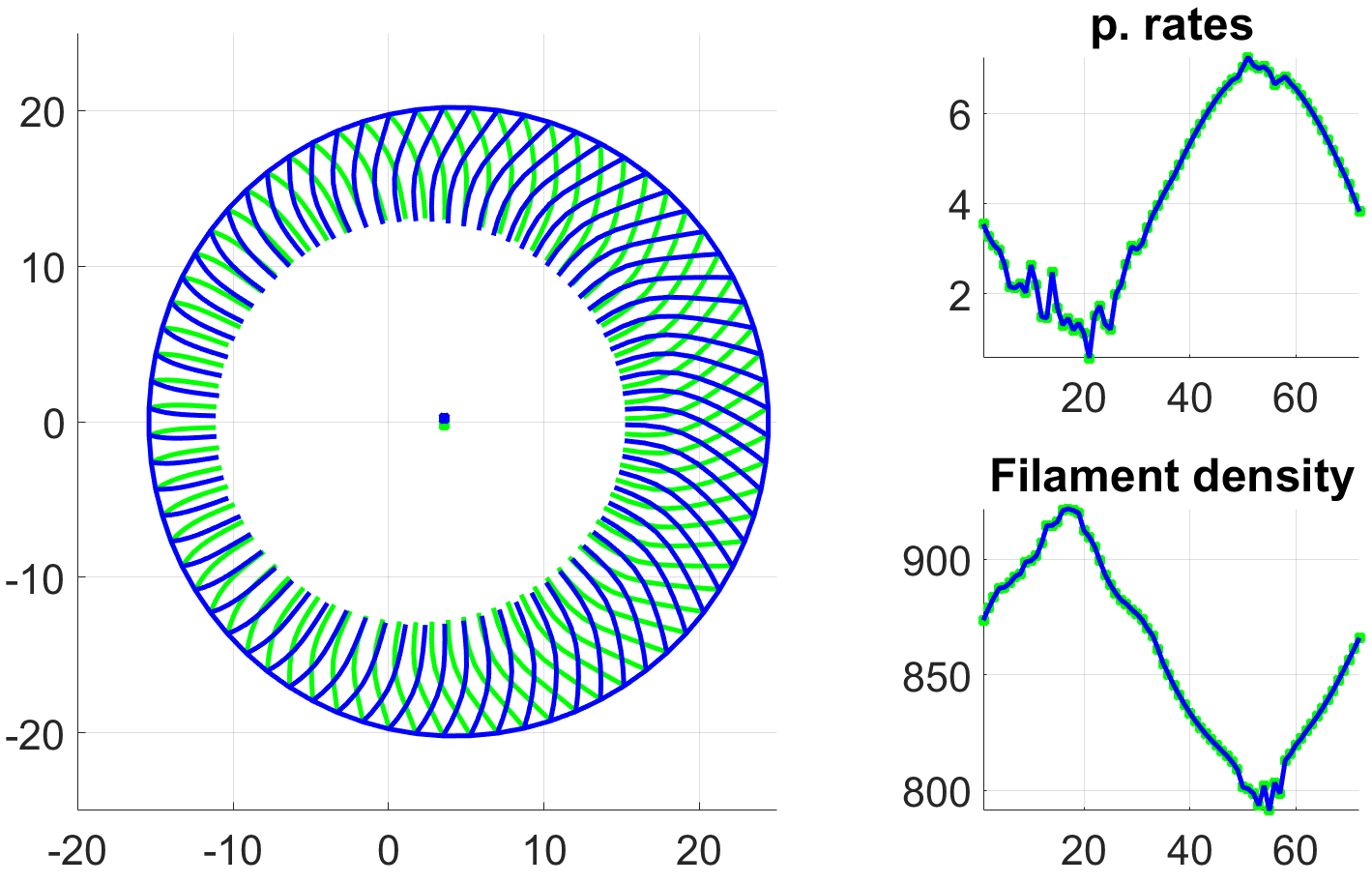}}\\
		\footnotesize{(a) $\Delta t=0.002$} & \footnotesize{(b) $\Delta t = 0.01$}
	\end{tabular}
	\caption{Experiment \ref{exp:dt.chemo} with $\snodes=9$, $\anodes=72$ and $\Delta t =0.002$ in (a) and $\Delta t =0.01$ in (b). Final time conformation. In both cases, the filaments in the lamellipodium are smooth but the polymerization rates in (b) are clearly not.} \label{fig:9x72}
\end{figure}

As a first step for the computation of $C$, we set the resolution of $B_0$ in \eqref{eq:cell} to be $\anodes=72$ and $\snodes=18$, along the $\a$ and $s$ directions respectively. Accordingly, \eqref{eq:CFL} recasts to $\Delta t \leq  0.017\,C$.  Using Experiment \ref{exp:dt.chemo} and varying $\Delta t$ by small increments  we identify two smooth conformations: one for $\Delta t=0.001$ and one for $\Delta t=0.0002$, cf. Figure \ref{fig:18x72}. When these timesteps are combined with \eqref{eq:CFL} they yield respectively
\begin{equation}\label{eq:C.choice}
	C=\frac{1}{15}\text{ and }C=\frac{1}{75}.
\end{equation}

To distinguish between these two values of $C$, we consider the coarser grid with $\snodes=9$ and $\anodes=72$ and we deduce from \eqref{eq:CFL} and \eqref{eq:C.choice} the timesteps $\Delta t = 0.01$ and $\Delta t = 0.002$ respectively.  We test these $\Delta t$ values in Experiment \ref{exp:dt.chemo} and note that in both cases the conformation of the lamellipodium is ``smooth'', cf. Figure \ref{fig:9x72}. In the case $\Delta t = 0.01$ though, the corresponding polymerization rate functions are non smooth; this implies timestep driven numerical instabilities in the solution. In effect, the value $C=\frac{1}{15}$ is too large, and hence we promote the value $C=\frac{1}{75}$.

\begin{figure}[t]
    \centering
	\begin{tabular}{cc}
%	\hskip -0.7em
		{\includegraphics[width=0.45\linewidth]{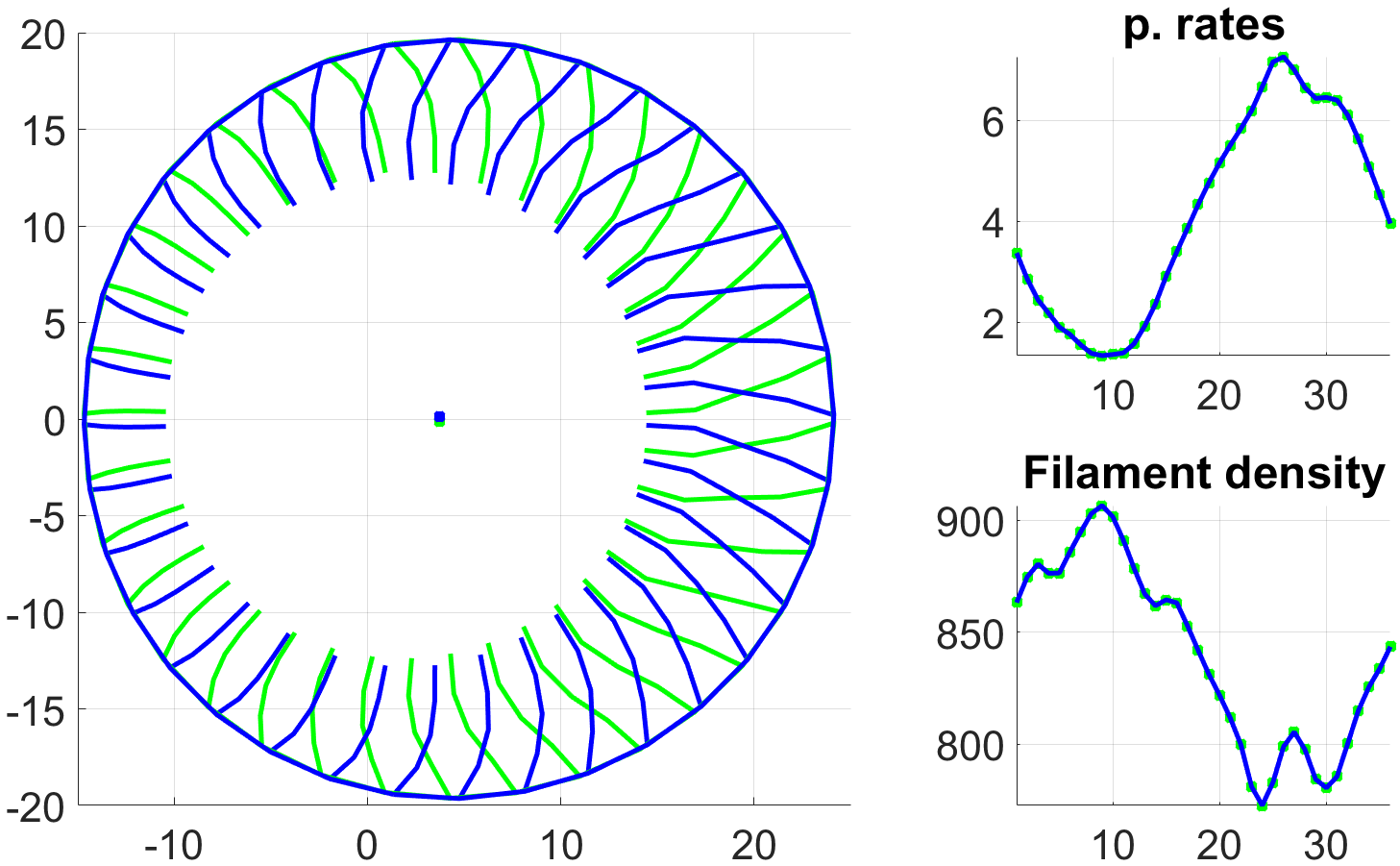}}
	   &{\includegraphics[width=0.45\linewidth]{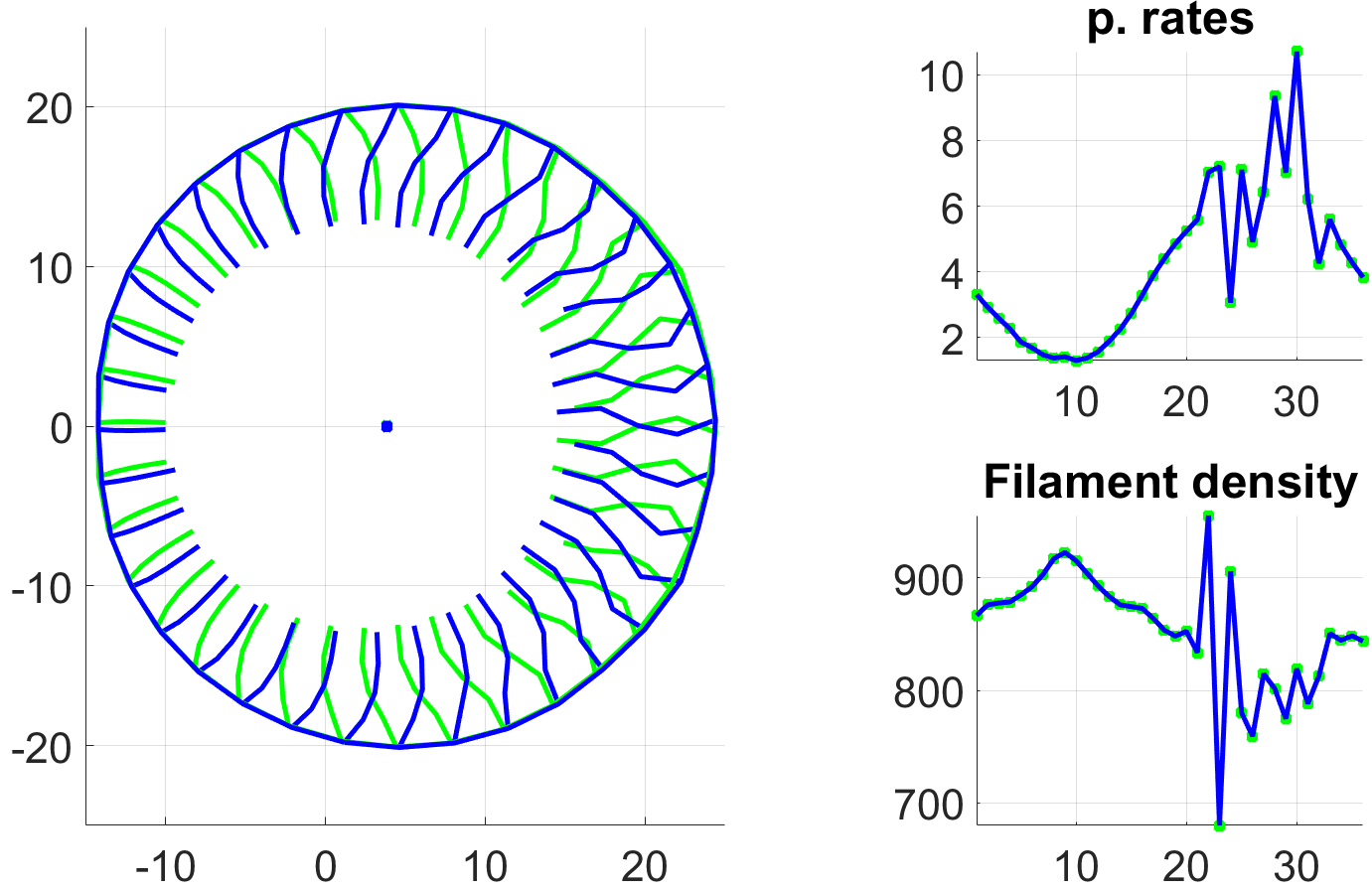}}\\
	   \footnotesize{(a) $\Delta t = 0.02$} & \footnotesize{(b) $\Delta t=0.08$}
	\end{tabular}
	\caption{Experiment \ref{exp:dt.chemo} with $\snodes = 5$, $\anodes = 36$. Final time. Both the filaments and the polymerization rate function in (b)  are not smooth; an indication that the corresponding timestep is too large.}\label{fig:5x36}
\end{figure}

We verify this stability constant with an even coarser grid with $\snodes = 5$ and $\anodes = 36$.  This time, we deduce from \eqref{eq:C.choice} and \eqref{eq:CFL}, the timesteps $\Delta t = 0.08$ and $\Delta t=0.02$ respectively. The simulation results are shown in Figure \ref{fig:5x36}, where we note that when $\Delta t = 0.08$ the filaments in the lamellipodium and the polymerization rates are not smooth. Hence, we promote once again the stability constant 
\begin{equation}\label{eq:C.chemo}
	C=\frac{1}{75}. 
\end{equation}

%In effect, the stability condition \eqref{eq:CFL} recasts into:
%\begin{equation}\label{eq:CFL_cond}
%	\Delta t \leq \frac{1}{75} {(\Delta s)^3\, \Delta\a}.
%\end{equation}

It is understood that this stability constant depends on a number of environmental parameters and variables, most notably on the extracellular chemical and matrix. Nevertheless, it is important to note that the FEM clearly exhibits stability regions with respect to the time and space steps $\Delta t$ and $\Delta \a$, $\Delta s$. 

\subsection*{Choosing the timestep}
What we have seen by the previous analysis is that the stability of the method can be identified by the smoothness of the polymerization rate and the filament density functions. Using this information we can adjust the timestep of the method in an automated way, similar to the \textit{adaptive timestep control} (ATC) methods, \cite{hundsdorfer2003numerical}. For that, we propose to use the \textit{divided difference smoothness measure} \cite{LeVeque2002,Sfakianakis.2016} or the $\beta_0$ \textit{smoothness indicators} used in \textit{weighted essentially non-oscillatory} (WENO) schemes, see e.g. \cite{Shu2009}. 

In some more detail, we consider the discrete polymerization rate (and/or the filament density) function $\left\{v_i,\, i=1\dots\anodes\right\}$ of the filaments at the end of one timestep of the method, see e.g. upper right panel in Figures \ref{fig:18x72}, \ref{fig:9x72}, \ref{fig:5x36}. 
	
%We compute the \textit{smoothness indicator}  as
%$$\theta_i =\frac{v_{i+1}-v_i}{v_i-v_{i-1}},\ i\in\{1\dots\anodes\},$$
%where the boundary cases $i=1$ and $i=\anodes$ are treated periodically.
%
%\remm{how about the sign at the maximum?}
	
The $\beta_i^0$ smoothness indicator is computed for $i\in {1\dots\anodes}$ as
\begin{equation}\label{eq:smooth}
	\beta_i^0 = \frac13 \(4v_{i-1}^2 - 13v_{i-1}v_i + 13v_i^2 + 5v_{i-1}v_{i+1} - 13v_iv_{i+1} + 4v_{i+1}^2\),
\end{equation}
where the ``boundary'' terms $\beta_1^0$ and $\beta_{\anodes}^0$ are computed periodically with respect to $i$, and where $\beta_i^0\geq 0$ (by construction). The formula \eqref{eq:smooth} is derived from a fourth order polynomial fitted to the numerical values $v_i$. This ``assumption'' is made to meet the smoothness of the FEM shape functions, i.e. third order along the $s$- and first order along the $\a$-direction, see \eqref{eq:H}, \eqref{eq:LG}.

The results ought to be understood as follows: the larger $\beta_i^0$ is, the ``less smooth'' the discrete function $\left\{v_i,\, i=1\dots\anodes\right\}$ is, see also \cite{Shu2009} for details. 

Accordingly, we update the timestep $\Delta t$ of the method using the rule 
\begin{equation}
	\Delta t^{\text{upd}} =\left\{\begin{aligned}
	&1.1\,\Delta t,&& \max_{i} \beta_i^0\leq \frac{2}{3}\,\beta_\text{thr}\\
	&0.9\,\Delta t,&& \max_{i} \beta_i^0> \frac{4}{3}\beta_\text{thr}\\
	&\Delta t,&& \text{else}
\end{aligned}\right.\,,
\end{equation}
where the threshold value $\beta_\text{thr}>0$ is set experimentally. In the first case, the updated $\Delta t^{\text{upd}}$ is employed in the next step of the method, whereas in the second case the same step is repeated with the updated $\Delta t^{\text{upd}}$ value. Subsequent repetitions of the same step of the method might be needed in the second case.

\subsection*{Haptotaxis and dependence on the gradient of the ECM}
\begin{figure}[t]
	\centering
	\begin{tabular}{ccc}
		~\includegraphics[height=9em]{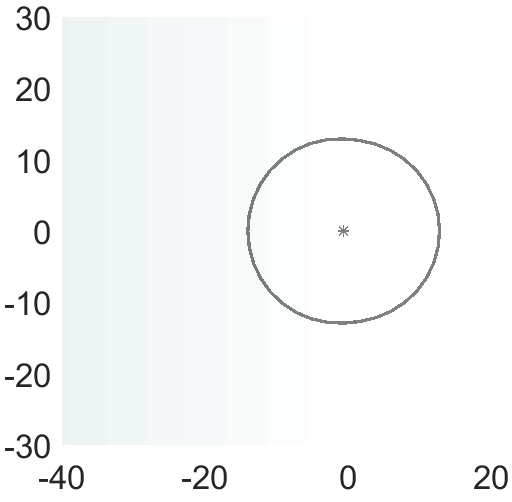}~
		&~\includegraphics[height=9em]{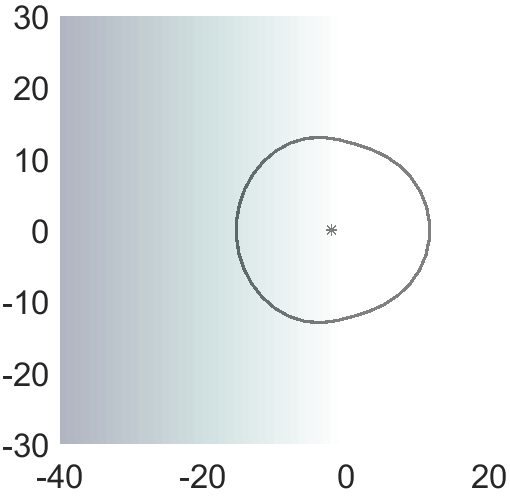}~
		&\includegraphics[height=9em]{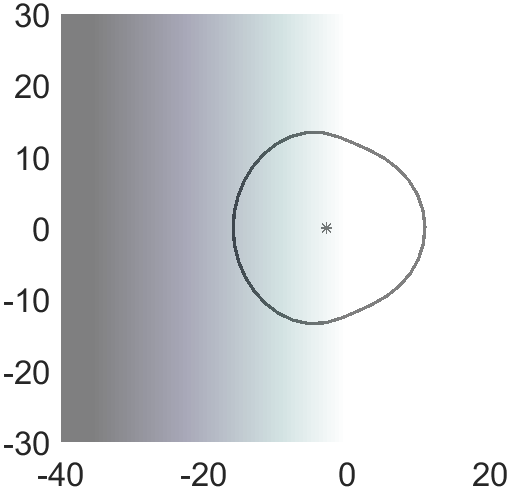}
		~\includegraphics[height=9em]{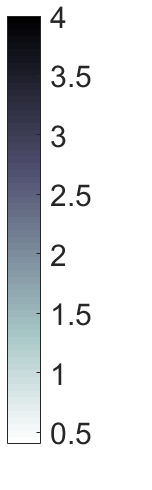}
		%\raisebox{-0.21em}{\includegraphics[height=9.38em]{embed_cbar_chem}}\\
		\\
		\footnotesize{(a) $\mu^A_1$}
		&\footnotesize{(b) $\mu^A_2$}
		&\footnotesize{(c) $\mu^A_3$}
	\end{tabular}		
	\caption{Experiment \ref{exp:dt.hapto} for the three different adhesion coefficients \eqref{eq:mu_A_1}--\eqref{eq:mu_A_3}. Showing here the final time simulations where it is clearly exhibited that the stronger the gradient of the ECM is, the more pronounced the deformation of the cell is. The colorbar on the right refers to the ECM and is common for all three figures.}\label{fig:dt.hapto}
\end{figure}

The second experiment that we consider in the time-step stability study is a haptotaxis one. This time though  we make one more step and identify the relation between the stability constant $C$ and the gradient of the ECM. We do so in the following experiment:

\begin{experiment}[Stability -- Haptotaxis]\label{exp:dt.hapto}
	An initially rotational symmetric cell lies over a non-uniform adhesion substrate. We consider three different cases that are incorporated in \eqref{eq:strong} by the adhesion coefficients:
	\begin{subequations}
	\begin{align}
		\mu^A_1(\vec x) &=\begin{cases}  0.4 - 0.01\,x, & x< 0\\ 0.4, & x\geq0 \end{cases},\label{eq:mu_A_1}\\
		\mu^A_2(\vec x) &=\begin{cases}  0.4 - 0.05\,x, & x< 0\\ 0.4, & x\geq0 \end{cases},\label{eq:mu_A_2}\\
		\mu^A_3(\vec x) &=\begin{cases}  0.4 - 0.1\,x, & x< 0\\ 0.4, & x\geq0 \end{cases}.\label{eq:mu_A_3}
	\end{align}	
	\end{subequations}
	where $\vec x=(x,y)\in \Omega=[-40,20]\times[-30,30]$. We assume that the chemical environment is uniform to a level that the polymerization rate is approximately\footnote{Small variations might emanate from the variable curvature of the membrane, see Section \ref{sec:model} and \cite{MOSS-model} for details.} $v^\pm(\a,t)=1$ for all filaments.  The domain is discretized with $\snodes = 7$ and $\anodes=36$, and the rest of the parameters are given in Table \ref{tbl:parameters} and the simulation results are shown in Figure \ref{fig:dt.hapto}. 
\end{experiment}

	As with Experiment \ref{exp:dt.chemo}, we identify for each $\mu^A$ case  \eqref{eq:mu_A_1}--\eqref{eq:mu_A_3}, the corresponding stability constant $C$ in \eqref{eq:CFL}. We do so by varying the timestep of the method by small increments and find the largest $\Delta t$ that still delivers smooth results.
	
\begin{figure}[t]
	\centering
	\footnotesize
	\begin{tabular}{c}
		{\includegraphics[width=0.4\linewidth]{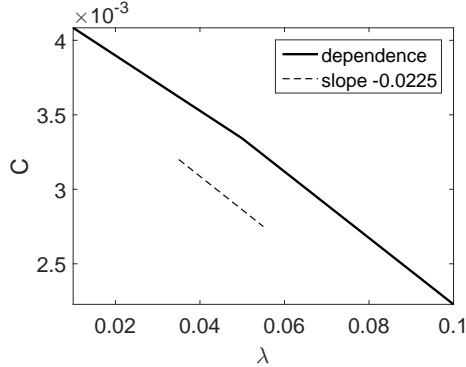}}
	\end{tabular}		
	\caption{Experiment \ref{exp:dt.hapto}. Showing the linear dependence of the stability constant $C$ on the ``gradient'' of the ECM $\lambda$, cf.  \eqref{eq:mu_A_1}--\eqref{eq:mu_A_3} and Figure \ref{fig:dt.hapto}.} \label{fig:dependence}
	\end{figure}

	This way, we identify the stability constants 
	\[
		C_1=0.0041\,,\quad  C_2= 0.0033\,,\quad  C_3=0.0022.
	\] 
	respectively for $\mu^A_1$, $\mu^A_2$, $\mu^A_3$. As the motility of the cell is haptotaxis driven, we correlate $C_1$ through $C_3$ with the corresponding (norm of the) gradients of the ECM
	\[
		\lambda_1=0.01\,,\quad \lambda_2=0.05\,,\quad \lambda_3=0.1.
	\]	
	The relation between the stability constant $C$ and the gradient $\lambda$ of the ECM is approximately linear 
	\begin{equation}\label{eq:C.hapt}
		C \sim -0.0225 \lambda, 
	\end{equation}
	see also Figure \ref{fig:dependence}. Clearly, the coefficient $-0.0225$ might depend on several cellular and extracellular parameters. Nevertheless, it is clearly exhibited by this experiment that the timestep stability constant depends in a linear decreasing way on the gradient of the ECM.

%	Summarising for this section, the numerical stability thresholds that we have identified in  Experiments \ref{exp:dt.chemo}, \ref{exp:dt.hapto} and \eqref{eq:C.chemo}, \eqref{eq:C.hapt} indicate a behaviour that is consistent with a \textit{numerically stable method}. In both cases the stability regions for the timestep $\Delta t$ are clearly defined and can be easily identified, even in an automated computational way.

%--------------------------------------------------------------------------	
%--------------------------------------------------------------------------
\section{Convergence of the FEM}\label{sec:converge}
As is typically done in complex models and methods, we exhibit here the convergence of the FEM in characteristic numerical experiments. In particular, we consider \change{a}{} chemotaxis, \change{a}{} haptotaxis, and \change{a}{} chemo-haptotaxis experiments. 

In each of these experiments we vary $\Delta s$ and/or $\Delta \a$ and compute cascades of numerical solutions over nested refined grids. As the exact solutions are not known, we compare every numerical solution to the one of the finest grid. 

The experiments that we consider are the following:

\begin{experiment}[Convergence -- Chemotaxis]\label{exp:conv:chemo}
	The cell migrates over a uniform adhesion substrate, under the influence of a chemical stimulus. The effect of chemotaxis is incorporated as a variable polymerization rate at the membrane. It varies smoothly $v_{\min}=1.5$ at the posterior side of the lamellipodium, to $v_{\max}=8$ at the anterior, see \cite{MOSS-model} for more details. The rest of the parameters are given in Table \ref{tbl:parameters}. 
\end{experiment}

\begin{experiment}[Convergence -- Haptotaxis]\label{exp:conv:hapt}
	The cell migrates within a chemically uniform environment over a non-uniform adhesion substrate. The variable ECM density is included in \eqref{eq:strong} by the adhesion coefficient,
	\[
		\mu^A(\vec x) = 0.4101\, \begin{cases} 0.1, & x<0\\ 0.1 + x/30, & x\geq 0 \end{cases},\quad \vec x =(x,y)\in \Omega 
	\]
	The polymerization rate is set at the same value\footnote{Small changes might occur due to variations in the curvature of the membrane, see Section \ref{sec:model} and \cite{MOSS-model} for details.} $v=8$ for all the filaments. The rest of the parameters are set as in Table \ref{tbl:parameters}.
\end{experiment}

\begin{figure}[t]
	\centering
	\footnotesize
	\begin{tabular}{cc}
		{\includegraphics[width=0.45\linewidth]{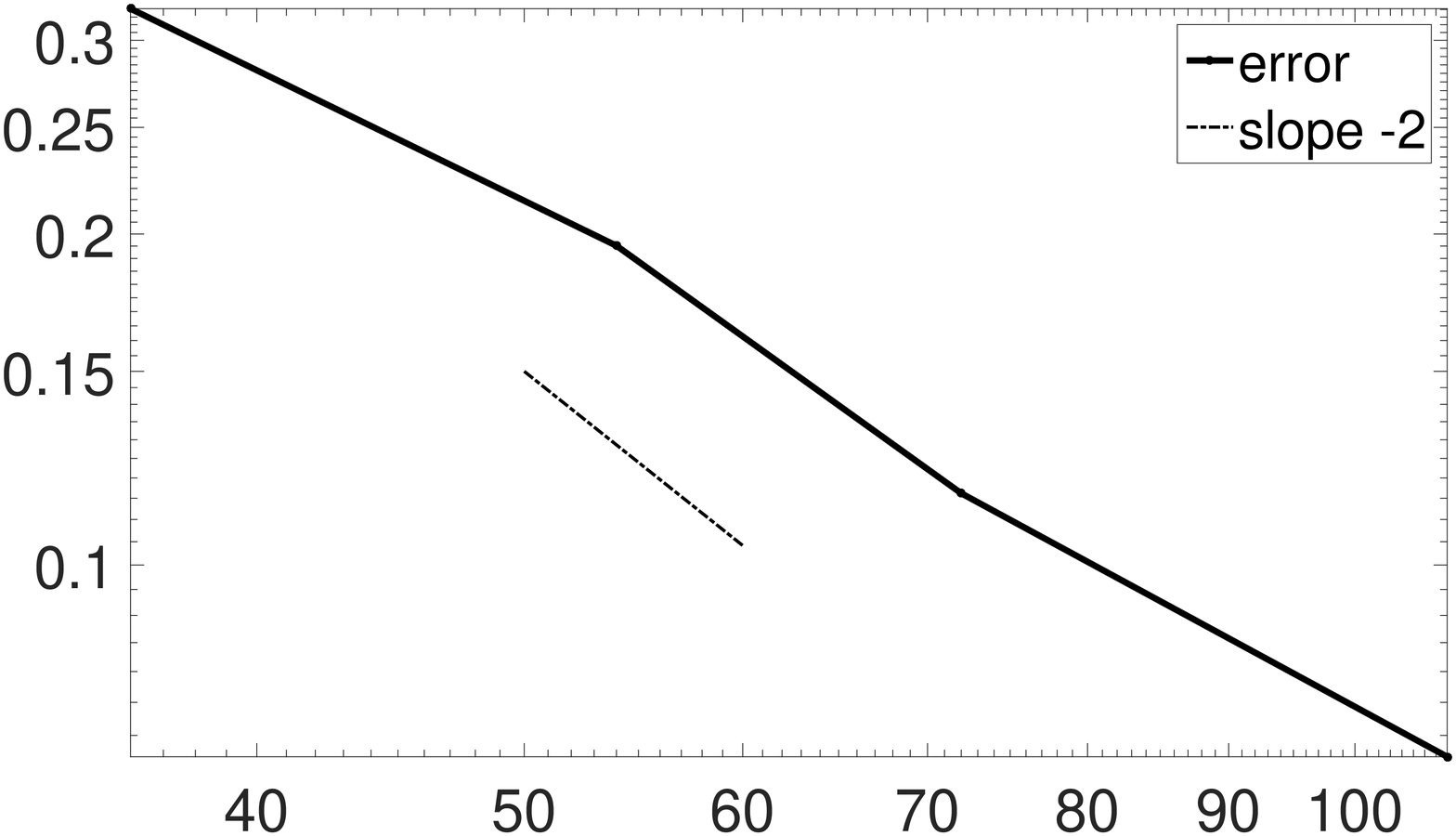}}
			&{\includegraphics[width=0.45\linewidth]{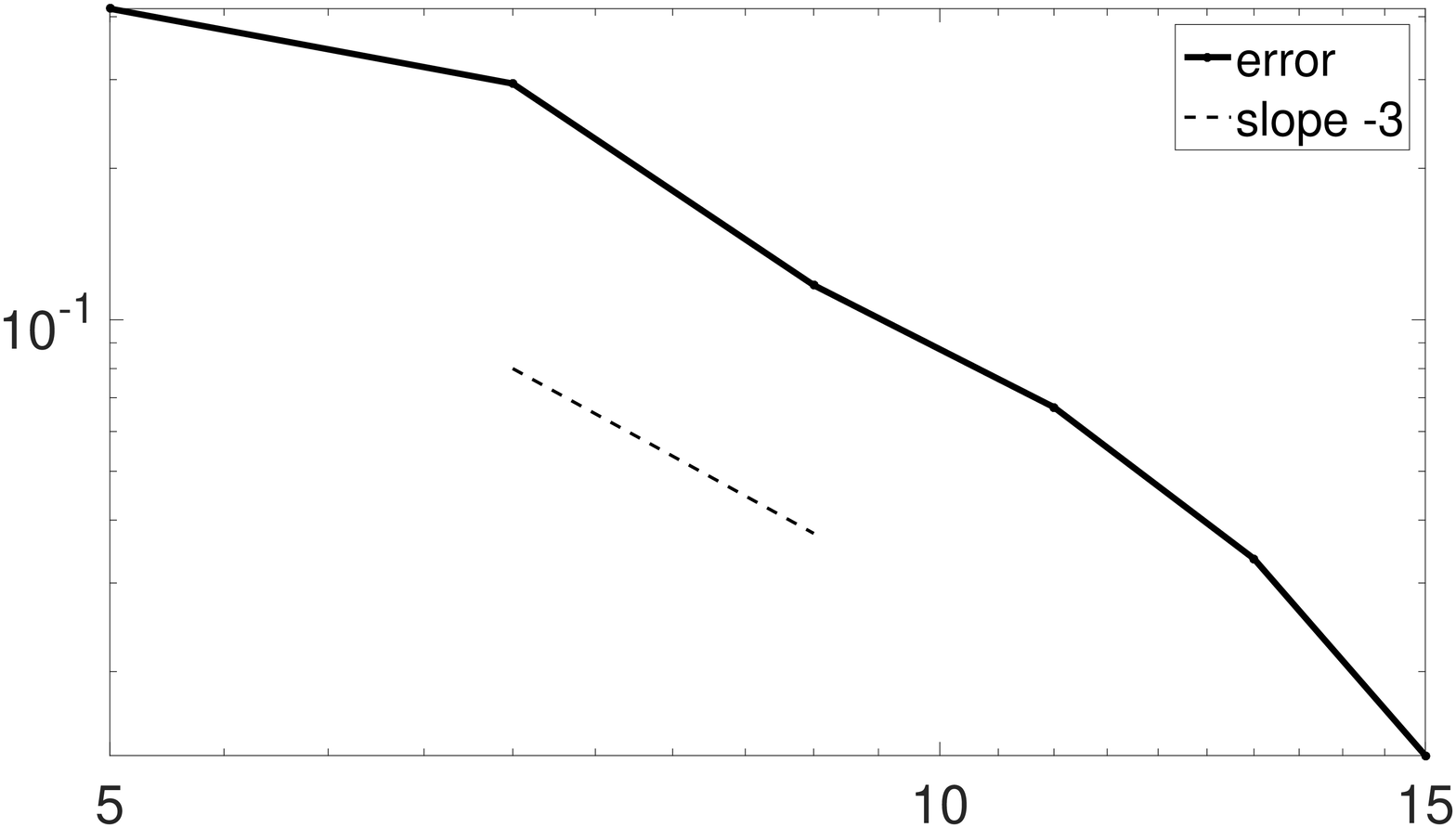}}\\[-0.7em]
		$\anodes$ & $\snodes$\\[0.3em]
		(a) $\snodes=9$ & (b) $\anodes=36$
	\end{tabular}		
	\caption{Convergence (log-log) plots for the Experiment \ref{exp:conv:chemo} (chemotaxis) using the ``norm'' $|\cdot|_A$. The ``error'' is computed as the difference against the numerical solution of the finest grid.  In {\rm(a)} we fix the value $\snodes=9$ and study the convergence with respect to $\anodes$ (horizontal axis). We compare the resulting convergence curve with the slope $-2$. In {\rm(b)} we fix $\anodes=36$ and study the convergence with respect to $\snodes$ (horizontal axis). The convergence rate in this experiment is comparable to 3.} \label{fig:conv1}
\end{figure}

\begin{experiment} [Convergence -- Chemo-haptotaxis]\label{exp:conv:mixed}
	The cell migrates under the influence of a chemical signal and over a non-uniform adhesion substrate. The polymerization rates and the adhesion coefficients are set as in Experiments \ref{exp:conv:chemo} and \ref{exp:conv:hapt}, i.e.
	\[
		\mu^A(\vec x) = 0.4101\, \begin{cases} 0.1, & x<0\\ 0.1 + x/30, & x\geq 0 \end{cases},\quad \vec x =(x,y)\in \Omega 
	\]
	and $v$ varies smoothly from $v_{\min}=1.5$ at the posterior side of the lamellipodium, to $v_{\max}=8$ at the anterior.
\end{experiment}

For each of the these experiments, the comparison between the results takes place in physical space. We denote by $\mathcal C\subset \mathbb{R}^2$ the full \textit{cell} (area enclosed by the outer border of the lamellipodium) consider the following ``norms'': 
\begin{subequations}
\begin{description}
	\item[Invasiveness:] The maximum $x$-coordinate of the cell: 
	\begin{equation}\label{eq:norm.invad}
		|\mathcal C|_A=\max_{(x,y)\in \mathcal C} x.
	\end{equation}
	This ``norm'' is more suited for experiments where the cell migrates to the right.
	
\item[Size:] The area of the minimum quadrilateral that the cell occupies:
	\begin{equation}\label{eq:norm.size}
		|\mathcal C|_B= \(\max_{(x,y)\in \mathcal C} x - \min_{(x,y)\in \mathcal C} x \) \(\max_{(x,y)\in \mathcal C} y - \min_{(x,y)\in \mathcal C} y \).
	\end{equation}
	
\item[Perimeter:] The length of the membrane of the cell, which coincides with the outer boundary of the lamellipodium:
	\begin{equation}\label{eq:norm.perim}
		|\mathcal C|_C= \int_{\partial \mathcal C} 1 .
	\end{equation}
	We compute this ``norm'' as the length of the closed \textit{piecewise-linear} curve defined by the outer ends of the discretization filaments. 
	
\item[Elongation:] The ratio of the sides of the minimum quadrilateral that the cell $\mathcal C$ occupies:
	\begin{equation}\label{eq:norm.elong}
		|\mathcal C|_D= \frac{\max_{(x,y)\in \mathcal C} x - \min_{(x,y)\in \mathcal C} x } { \max_{(x,y)\in \mathcal C} y - \min_{(x,y)\in \mathcal C} y}.
	\end{equation}
\end{description}
\end{subequations}

\begin{figure}[t]
	\centering
	\footnotesize
	\begin{tabular}{cc}
		{\includegraphics[width=0.45\linewidth]{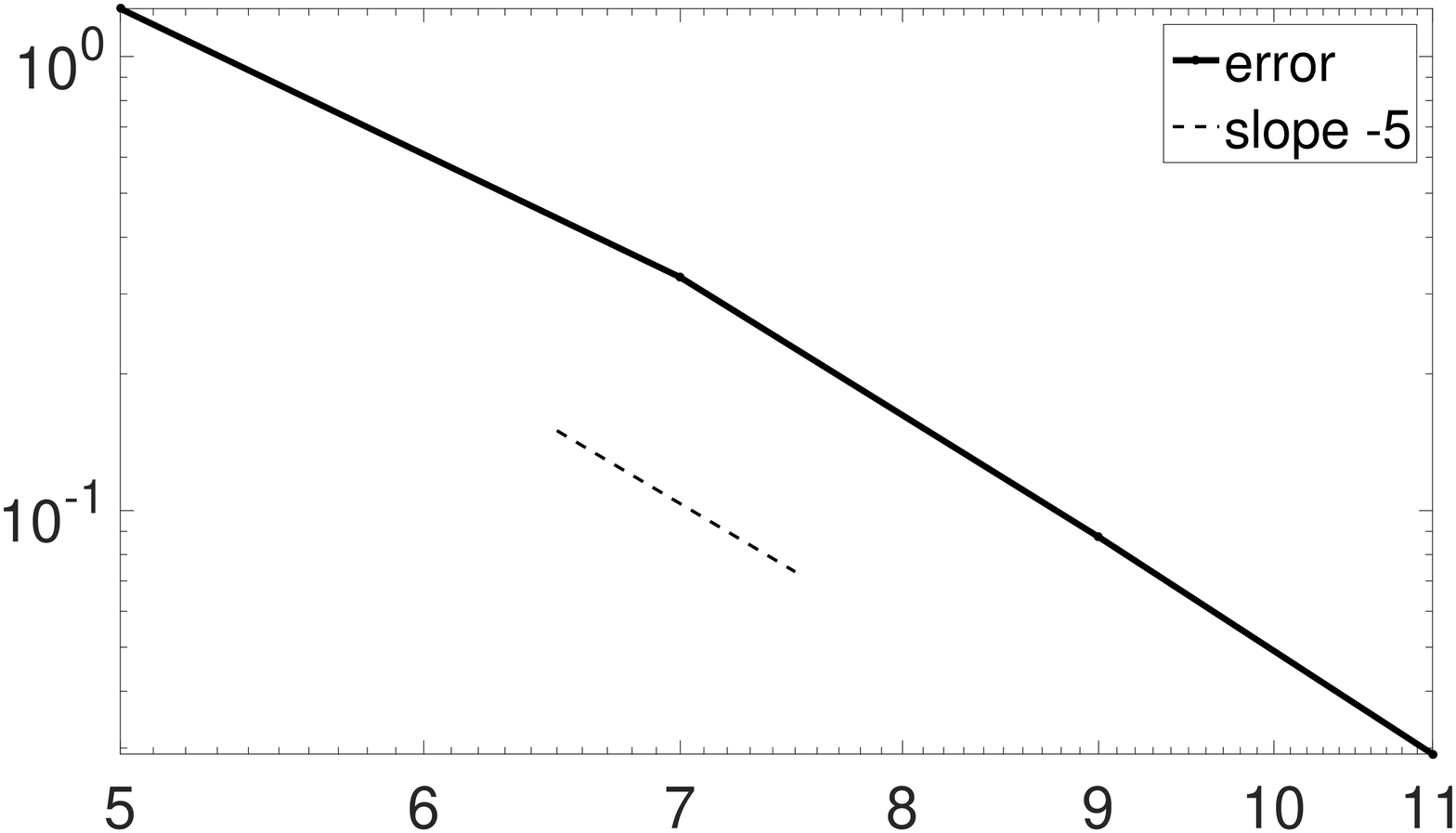}}
		&{\includegraphics[width=0.46\linewidth]{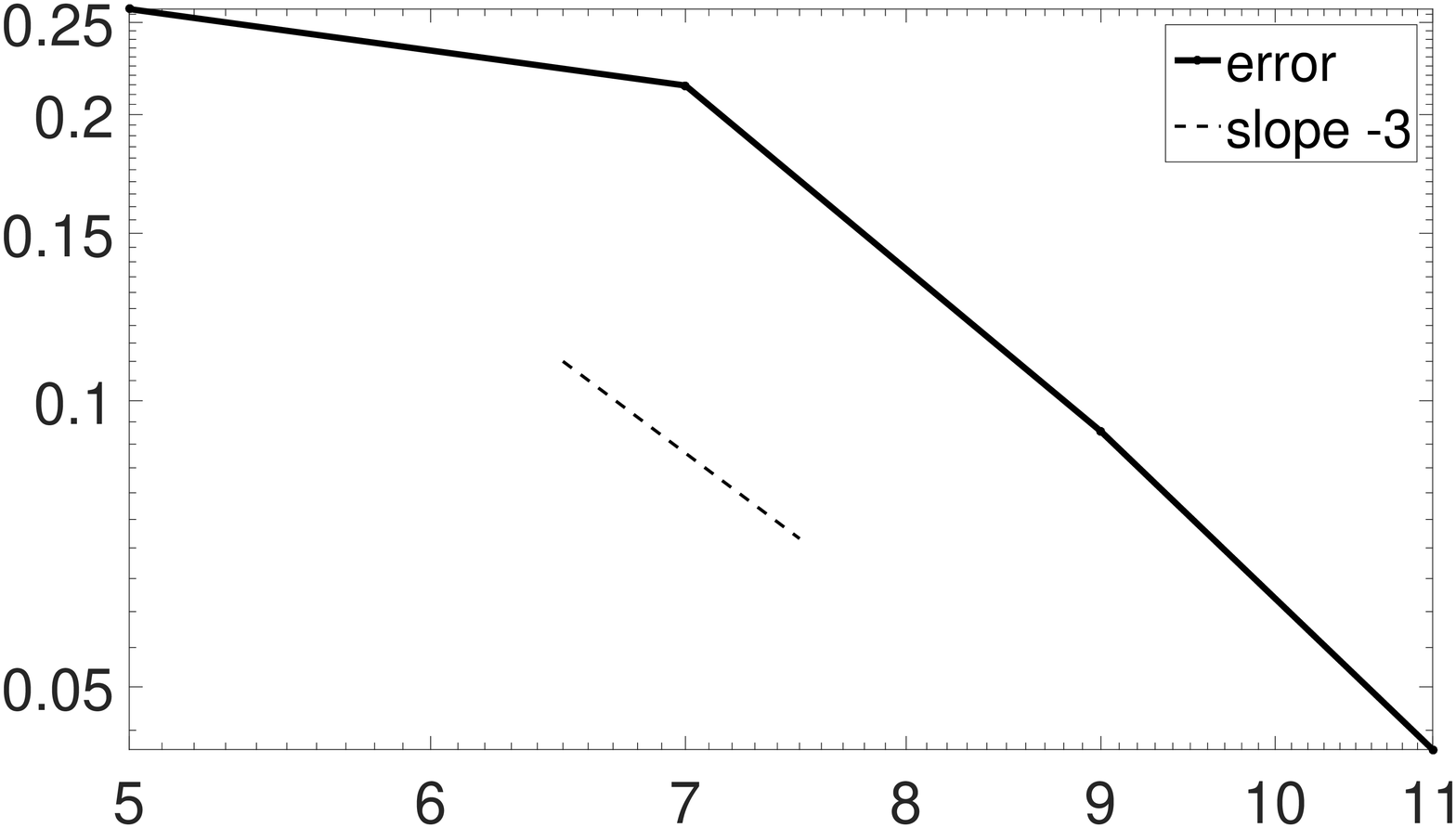}}\\[-0.7em]
		$\snodes$ & $\snodes$\\[0.3em]
		(a) haptotaxis, $\anodes=36$ & (b) chemo-haptotaxis, $\anodes=144$
	\end{tabular}		
	\caption{Convergence (log-log) plots in the haptotaxis and the combined chemo-haptotaxis Experiments \ref{exp:conv:hapt}, \ref{exp:conv:mixed} with respect to the the $|\cdot|_A$  ``norm''. In {\rm(a)} we consider the haptotaxis Experiment \ref{exp:conv:hapt} and study the convergence with respect to $\snodes$ (horizontal axis) for fixed $\anodes=36$. In {\rm(b)} we consider the chemo-haptotaxis Experiment \ref{exp:conv:mixed} and study the convergence with respect to $\snodes$ (horizontal axis) for fixed $\anodes=144$.}\label{fig:conv2}
\end{figure}

\begin{figure}[t]
	\centering
	\footnotesize
	\begin{tabular}{cc}
		\includegraphics[width=0.45\linewidth]{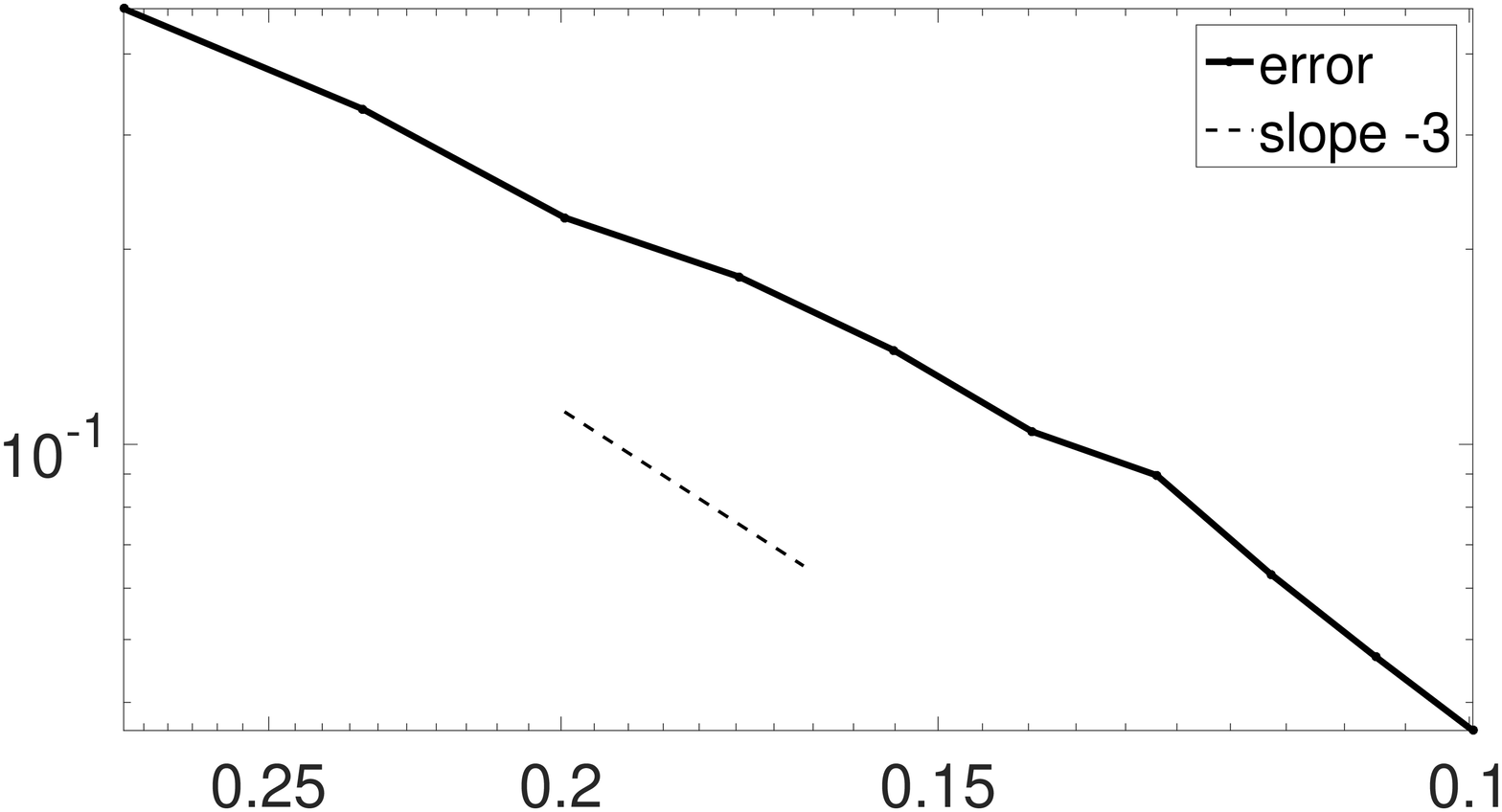}
		& \includegraphics[width=0.45\linewidth]{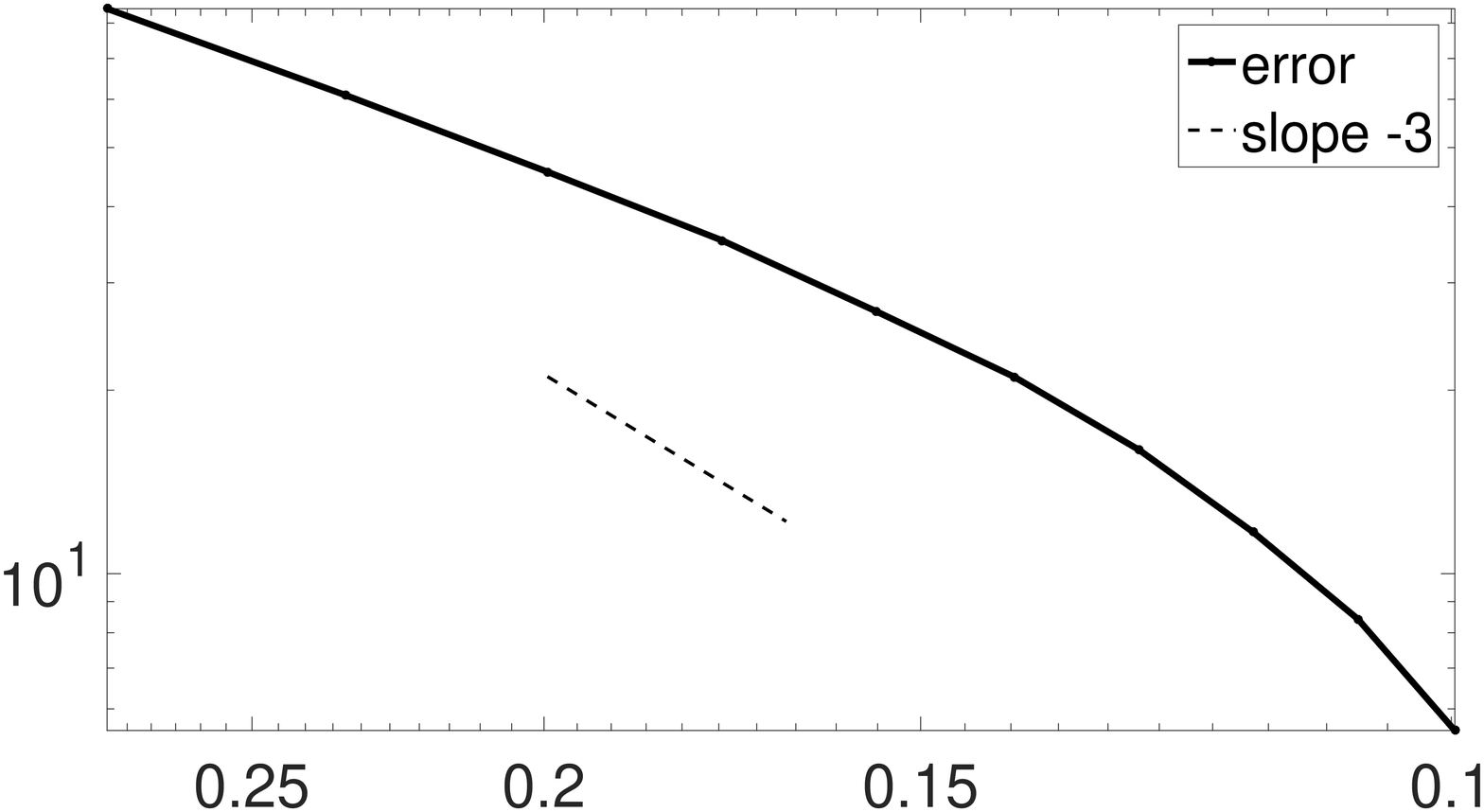}\\[-0.4em]
		~~~~~inscribed circle diameter &~~~~~inscribed circle diameter\\[0.3em]
		(a) norm $|\cdot |_A$ & (b) norm $|\cdot |_B$\\
	\end{tabular}		
	\caption{Convergence (log-log) plots for the chemo-haptotaxis Experiment \ref{exp:conv:mixed} using the $|\cdot|_A$ and the $|\cdot|_B$ norms. In both (a) and (b),  $\anodes$ and $\snodes$ vary at the same time. The convergence is studied with respect to the (decreasing) radius of the inscribed circle (horizontal axis) of the discretization cells $C_{i,j}$ \eqref{eq:cell}. The convergence rates are comparable to 3.}\label{fig:conv3}
\end{figure}

We first define the \textit{distance} between two cells $\mathcal C_1,\ \mathcal C_2$ as 
\begin{equation}\label{eq:diff.cells}
	d_X(\mathcal C_1,\mathcal C_2) = \big| |\mathcal C_1|_X - |\mathcal C_2|_X \big|,
\end{equation}
where $|\cdot|_X$, $X=A,\,B,\,C$, or $D$, represents the ``norm'' of choice. The convergence results with respect to the $|\cdot |_A$ and $|\cdot |_B$ ``norms'' can be found in Figures \ref{fig:conv1} -- \ref{fig:conv3}; similar results are obtain for the $|\cdot |_C$ and $|\cdot |_D$ ``norms''.

Figure \ref{fig:conv1} refers to the chemotaxis Experiment \ref{exp:conv:chemo}. We set the fixed value $\snodes=9$ and allow for $\anodes\in\left\{36,54,72,108\right\}$ to vary. For every instance of $\anodes$ we compute the ``error'' as the distance of the corresponding $\mathcal C_{\anodes}$ with the one with the finest grid, i.e. 
$$d_X(\mathcal C_{\anodes}, \mathcal C_{108}).$$
In this case, we deduce a convergence rate of about two. In a similar way, we consider the fixed value $\anodes=36$ and vary $\snodes\in\left\{5,7,\dots,15\right\}$ to deduce a convergence rate of approximately three. 

In Figure \ref{fig:conv2} we study the convergence of the haptotaxis Experiment \ref{exp:conv:hapt} and the chemo-haptotaxis Experiment \ref{exp:conv:mixed}. We set the fixed values $\anodes=36$ and $\anodes=144$ respectively, and allow for $\snodes\in\{5,7,9,11\}$ to vary. We once again deduce the convergence of the numerical method. When comparing with the convergence in the chemotaxis Experiment \ref{exp:conv:chemo} (shown in Figure \ref{fig:conv1}), the results obtained here indicate that the influence of the non-uniform ECM on the numerical error dissipates fast with respect to the resolution of the grid. In other words, coarse discretization grids are suited better for haptotaxis rather than for chemotaxis experiments.

In Figure \ref{fig:conv3} we consider again the mixed chemo-haptotaxis Experiment \ref{exp:conv:mixed}. This time though, the discretization varies with respect to both $\snodes$ and $\anodes$ at the same time. As the grid is refined, we consider the \textit{diameter of the inscribed circle of the discretization cell} $C_{i,j}$ as the controlling parameter, and study the convergence of the method with respect to it. We use the norms $|\cdot|_A$ and $|\cdot |_B$, and compare every numerical solution with the one of the finest grid. The detailed results for the $|\cdot|_A$ ``norm'' can be found in Table \ref{tbl:conv.mixed}, where the convergence rates are computed by the slope of piecewise linear curve.

%The time stability of the FEM in Section \ref{sec:stbility}, along with its convergence in Section \ref{sec:converge}, indicate that the rigorous numerical analysis of the FEM is warranted. 

%--------------------------------------------------------------------------	
%--------------------------------------------------------------------------
\section{Embedding the FBLM in the extracellular environment}\label{sec:env}
In this section we model and simulate the interactions between the FBLM-FEM and the extracellular environment. We start with a model that describes the extracellular environment and the cell, and with a short description of the numerical method that we use to solve it. We conclude this section with two particular numerical experiments that exhibit the combination of the FBLM-FEM with the environment.

%------------------------------------------------------------------------------------------------------------------
\subsection{A model for the environment}
We assume that the extracellular environment is comprised of the ECM, a chemical ingredient that serves as \textit{chemoattractant} for the cell, and \textit{matrix metalloproteinases} (MMPs) that are secreted by the cell and are responsible for the degradation of the matrix. These environmental components participate in our study via the density of the corresponding (macro-)molecules.

The extracellular chemical, denoted by $c$, is injected in the environment by one or more \textit{micro-pipettes} that are modelled here as source terms. The chemical is assumed to diffuse freely in the environment, to decay with time (chemical degradation), and to be degraded by the cell upon attachment. The MMPs, denoted by $m$, are produced by the cell, they diffuse freely in the environment, and decay with time. The ECM, denoted by $v$, is assumed to be an immovable component of the system that decays upon attachment of the MMPs, and is not remodelled. Overall the model of the environment reads
\begin{equation}\label{eq:env}
\left\{\begin{aligned}
\frac{\partial c}{\partial t}(\vec x,t) &= D_{c} \Delta c(\vec x,t)  + \sum_{i=1}^{N^\text{pip}}\a_i\, \mathcal X_{\mathcal P_i(t)}(\vec x)  - \gamma_1 c(\vec x,t) - \delta_1\, \mathcal X_{\mathcal C(t)}(\vec x)\\
\frac{\partial m}{\partial t}(\vec x,t) &= D_m \Delta m(\vec x,t) + \beta \mathcal X_{\mathcal C(t)}(\vec x) - \gamma_2 m(\vec x,t)\\
\frac{\partial v}{\partial t}(\vec x,t) &= -\delta_2 m(\vec x,t) v(\vec x,t)
\end{aligned}\right.
\end{equation}
with $\vec x\in \Omega\subset \mathbb R^2$, $t\geq 0$, and $D_c,\, D_m, \a_i, \beta, \gamma_i, \delta_i \geq 0$. The number of pipettes is $N^\text{pip}$ and the corresponding domain/source has support  $\mathcal P_i$, $i=1\dots N^\text{pip}$. The feedback of the FBLM-FEM to the environment takes place \change{at}{through} the term $\mathcal X_{\mathcal C(t)}(\vec x)$, where $\mathcal C(t)\subset \mathbb R^2$ represents the \textit{full cell} (lamellipodium and internal structures). 

\begin{table}[t]
	\footnotesize
	\begin{center}
		\begin{tabular}{c|c|c}
			inscribed circle & $|\cdot|_A$ & convergence \\
			diameter&``error''&rate\\
			\hline
			\\[-0.7em]
			0.27925 & 0.46966 & ---\\
			0.23271 & 0.32897 & 1.9529\\
			0.19947 & 0.22370 & 2.5022\\
			0.17453 & 0.18117 & 1.5788\\
			0.15514 & 0.13964 & 2.2108\\
			0.13963 & 0.10468 & 2.7357\\
			0.12693 & 0.08957 & 1.6343\\
			0.11636 & 0.06306 & 4.0358\\
			0.10740 & 0.04702 & 3.6627\\
			0.09973 & 0.03610 & 3.5346
		\end{tabular}
	\end{center}
	\caption{The data corresponding to the convergence Figure \ref{fig:conv3} (a) and the Experiment \ref{exp:conv:mixed} for the $|\cdot|_A$ ``norm''. The ``error'' of each numerical solution is computed by its difference against the numerical solution of the finest grid using \eqref{eq:diff.cells}. The convergence rates are computed by the slope of the corresponding segment of the curve.}\label{tbl:conv.mixed}
\end{table}

Clearly, \change{the}{} model \eqref{eq:env} is simple and accounts only for some of the basic extracellular processes and interactions between the FBLM and the environment. It can easily be extended to incorporate further and more precise biological properties/phenomena. As it is not though the main aim of this paper, we refrain from such generalizations here and postpone this study for a follow-up work.

As it is easier for the presentation of the numerical method, we write the system \eqref{eq:env} in an alternative operator form:
\begin{equation}\label{eq:env.oper}
\partial_t \vec w= R(\vec w) + D(\vec w)\;,
\end{equation}
where $\vec w=(c, m, v)^T$ and where $R$, $D$ are the reaction and diffusion operators respectively:
\begin{align}
D(\vec w)&= \(D_c \Delta c,\ D_m \Delta m,\ 0\)^T,\\
R(\vec w)& = \(\sum_{i=1}^{N^\text{pip}}\a_i \mathcal X_{\mathcal P_i}  - \gamma_1 c - \delta_1 \mathcal X_{\mathcal C},\ \beta \mathcal X_{\mathcal C} - \gamma_2 m,\ -\delta_2 m v\)^T.
\end{align}

The system \eqref{eq:env} (or \eqref{eq:env.oper}) is equipped with initial and boundary conditions, and parameters that are experiment specific.

For the numerical treatment of \eqref{eq:env.oper} we use a second order \textit{Implicit-Explicit Runge-Kutta} (IMEX-RK) \textit{Finite Volume} (FV) numerical method that was previously developed in \cite{Sfakianakis.2016, Sfakianakis.2016b} where we refer for more details. Here, in Appendix \ref{sec:FV}, we give some details.

%------------------------------------------------------------------------------------------------------------------
\subsection{Coupling the FBLM with the environment}
\begin{figure}[t]
	\centering
	\begin{tabular}{ccc}
		\raisebox{-0.21em}{\includegraphics[height=9.5em]{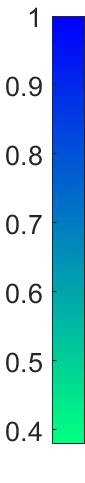}}~
		\includegraphics[height=9.3em]{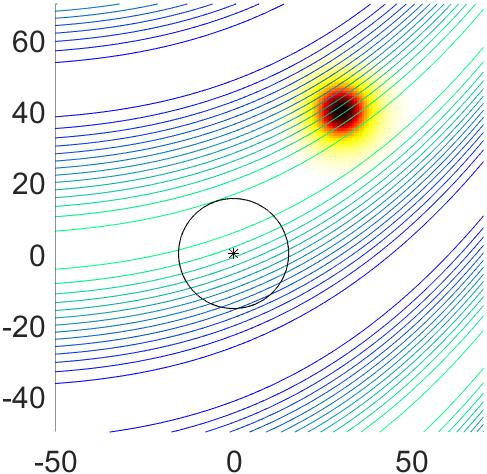}
		&\includegraphics[height=9.3em]{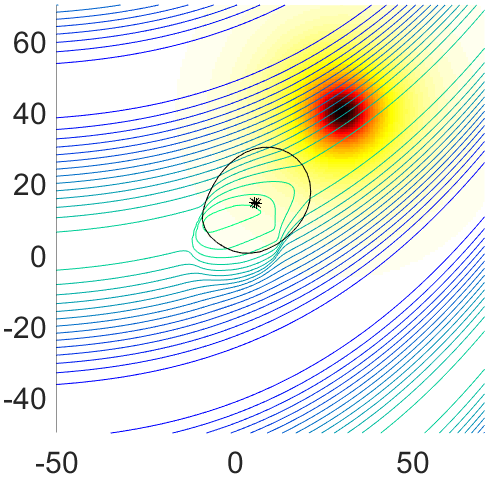}
		&\includegraphics[height=9.3em]{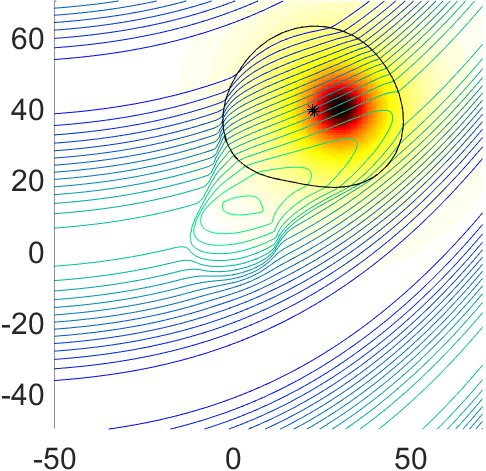}
		%		&\includegraphics[height=9em]{embed_cbar_ecm}
		~\raisebox{-0.21em}{\includegraphics[height=9.48em]{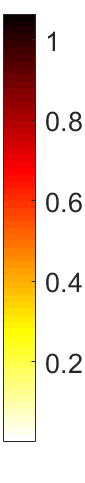}}\\
		\footnotesize{(a) $t=0.021$}
		&\footnotesize{(b) $t=10.021$}
		&\footnotesize{(c) $t=20.021$}
	\end{tabular}		
	\caption{Experiment \ref{exp:embedding} (In the environment): A cell (closed curve) migrates on a non-uniform ECM (isolines) under the influence of an extracellular chemical attractor injected in the environment by a circular pipette (upper right). {\rm (a):} As the chemical diffuses in the environment and decays, it creates a chemical gradient. {\rm (b)-(c):} As the cell identifies the gradients of the chemical and of the ECM, it responds by adjusting its motility and shape. The size of the cell increases with the intensity of the surrounding chemical. At the same time, the MMPs produced by the cell degrade the ECM. The colorbars refer to the densities of the ECM (left) and the chemical (right).}\label{fig:embedding}
\end{figure}
The FBLM-FEM \eqref{eq:strong} and the environment \eqref{eq:env} are coupled at three different places \change{}{through}: a) \change{in}{}the characteristic function $\mathcal X_{\mathcal C}$ in the model of the environment \eqref{eq:env}, where the cell $\mathcal C$ produces MMPs and degrades the chemical, b) \change{in}{}the adhesion coefficient $\mu^A$ of the FBLM in \eqref{eq:strong} that reflects the density of the ECM, see e.g. Experiment \ref{exp:conv:hapt}, and c) \change{in}{}the polymerization rate $v^\pm_\text{ext}$ in  \eqref{eq:chem.pol} which is primarily controlled by the intensity of the extracellular chemical $c$ at the membrane of the cell.

Numerically, the FEM and FV methods that solve the FBLM \eqref{eq:strong} and environment \eqref{eq:env} model, are combined in a modular way. During the time period $[t^n,t^{n+1}]$, $t^{n+1}=t^n+\Delta t^n$, $n=0, \dots$ the following hold
\begin{itemize}
	\item[---] The timestep $\Delta t^n$ is common in the FEM and the FV and is dictated by the stability of both methods, see also Section \eqref{sec:stbility}. 
	\item[---] The FV takes into account the position of the cell at $t^n$, i.e. the FV is explicit with respect to the cell.
	\item[---] Similarly, the FEM is explicit with respect to the density of the ECM and the chemical.
\end{itemize}

We exhibit the coupling of the cell with the environment, and their interactions with two particular experiments:
\begin{experiment}[In the environment]\label{exp:embedding}
	We consider an initial non-uniform adhesion substrate, a circular pipette that injects chemical in the environment, and a rotational symmetric cell in some distance from the pipette.
	The ECM is given by
	\begin{equation}
		\mu^A(\vec x) = 0.5\( \sin\( \(2y-x^3\)\pi \)^2 + 1\)\,, 	
	\end{equation}
	where $\vec x=(x,y)\in \Omega = [-50,60]\times [-50,70]$.
	
	The simulation results are shown in Figure \ref{fig:embedding}. The parameters for the FBLM and the environment models are given in Tables \ref{tbl:parameters} and \ref{tbl:env.parameters}.
	\end{experiment}
The chemical diffuses in the environment and is identified by the cell, which responds with a combined motion to its gradient and to the gradient of the ECM. While the cell migrates, it secretes MMPs that diffuse in the environment, decay, and degrade the ECM.
	
\change{The }{}Experiment \ref{exp:embedding} exhibits the relation between the size of the cell, the width of the lamellipodium, the polymerization rates of the filaments, and the density of the extracellular chemical attractant. We clearly see how the size of the cell increases as it approaches the pipette and the higher density of the chemical, cf. \eqref{eq:width_and_v} and \cite{MOSS-model,Brunk2016}.

\begin{figure}[t]
	\centering
	\begin{tabular}{cccc}
		\hspace{-0.2em}\includegraphics[height=7.2em]{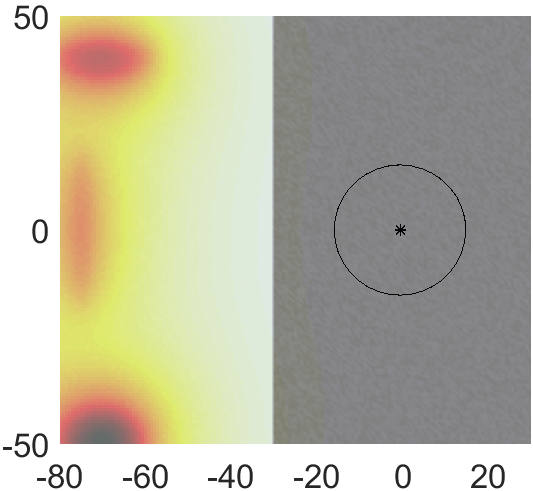}
		&\hspace{-0.3em}\includegraphics[height=7.4em]{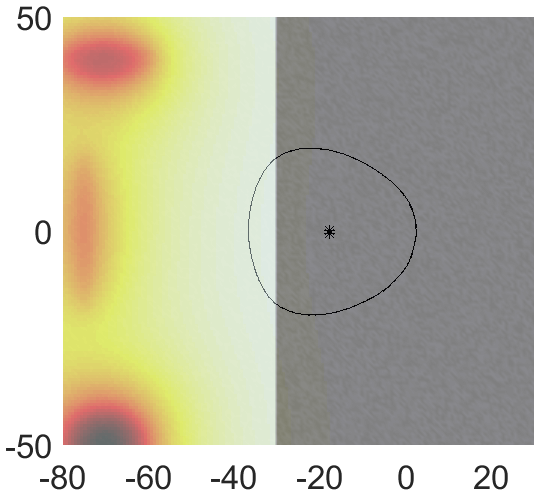}
		&\hspace{-0.3em}\includegraphics[height=7.4em]{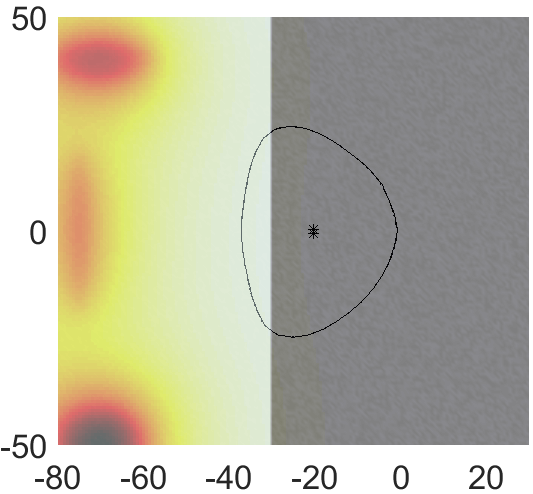}
		&\hspace{-0.2em}\includegraphics[height=7.4em]{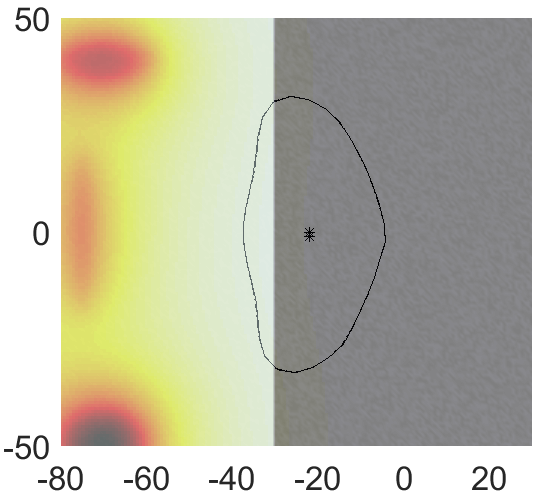}
		\includegraphics[height=7.1em]{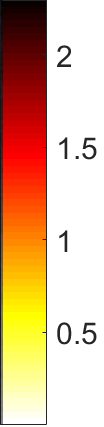}\\
		\footnotesize{(a) $t=0.571$}
		&\footnotesize{(b) $t=16.201$}
		&\footnotesize{(e) $t=20.611$}
		&\footnotesize{(e) $t=28.441$}
		%&\footnotesize{(d) cbar }%chemical and matrix}
	\end{tabular}		
	\caption{Time evolution of the Experiment \ref{exp:softstiff} (Adhesion wall): A cell faces a discontinuity in the ECM with density that is lower in the left side of the domain and higher density of adhesion sites perturbed by a small amount of noise. The colorbar on the right refers to the intensity of the chemical. {\rm(a):} Initially the cell is rotational symmetric and resides on the higher density part of the ECM. The three pipettes are located in the lower density part of the ECM and create a complex chemical environment that is attractive for the cell. {\rm(b-c):} The cell identifies the gradient of the chemical and migrates towards the pipettes. When it reaches the discontinuity of the ECM, it ceases further migration as it cannot exert sufficient adhesions. The part though that still resides on the higher density ECM, keeps on migrating and as an effect the cell elongates parallel \change{}{to} the discontinuity of the ECM.}\label{fig:softstiff}
\end{figure}

For the next experiment we are motivated by \cite{Chaplain2012, Lo2000}, and in particular by the opposing effects that chemical attraction and the lack of sufficient adhesion can have on the migration of the cell. As the aim of this paper is not to reproduce experimental scenarios, we postpone this more detailed work for a follow-up work. Nevertheless we present it here as an indication of the combination of the FBLM \eqref{eq:strong} and the environment \eqref{eq:env} with multiple chemical sources.

\begin{experiment}[Adhesion wall]\label{exp:softstiff}
	We consider three sources that inject the same chemical in the environment in a constant rate. The chemical diffuses and decays and a complex chemical landscape is formed in the environment. The ECM exhibits a jump-discontinuity between a higher and lower density that separates the domain in two parts. 
	\begin{equation}
		\mu^A(\vec x) = 0.5\,\begin{cases} 1, & x<-30\\ 5, & x\geq -30 \end{cases},\quad \vec x =(x,y)\in \Omega
	\end{equation}
	The three pipettes reside on the lower-ECM-density part of the domain ($x< -30$) whereas an (initially rotational symmetric) cell on the higher-ECM part ($x\geq -30$). 
	
	The simulation results of this experiment are given in Figure \ref{fig:softstiff} and the parameters \change{follow}{are as shown in} the Tables \ref{tbl:parameters} and \ref{tbl:env.parameters}.
\end{experiment}

As the chemical gradient is formed, it is identified by the cell which starts migrating towards the pipettes. The propagating front of the cell ceases further migration when the cell arrives at the discontinuity of the ECM, as it cannot create sufficient adhesions to the lower-density ECM and transfer its momentum. The rest of the cell, that resides on the higher-density ECM, keeps on migrating and the cell effectively elongates in a direction parallel to the discontinuity.

%--------------------------------------------------------------------------	
%--------------------------------------------------------------------------
\section{Sensitivity analysis}\label{sec:sensit}

We investigate here the sensitivity of the FBLM-FEM on several of its parameters. We consider a particular experiment and a reference parameter set and compute with them a reference numerical solution.  We then vary the parameter set, compute the new results, compare them with the reference solution, and quantify the effect the varied parameters have on the numerical solution.

As opposed to the study of the stability and the convergence in Sections \ref{sec:stbility} and \ref{sec:converge}, the sensitivity includes more biological information and meaning. For that reason, we employ now the FBLM augmented with the model for the environment \eqref{eq:env}.

The control experiment that we consider is a chemotaxis scenario with two sources of chemical.
 
\begin{figure}[t]
	\centering
	\begin{tabular}{ccl}
		\includegraphics[height=10em]{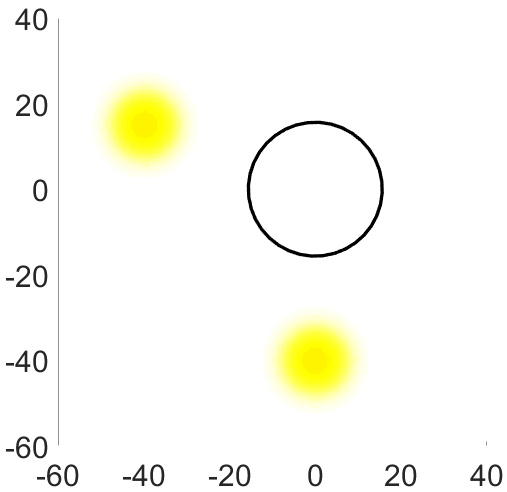}~~~
		& \includegraphics[height=10em]{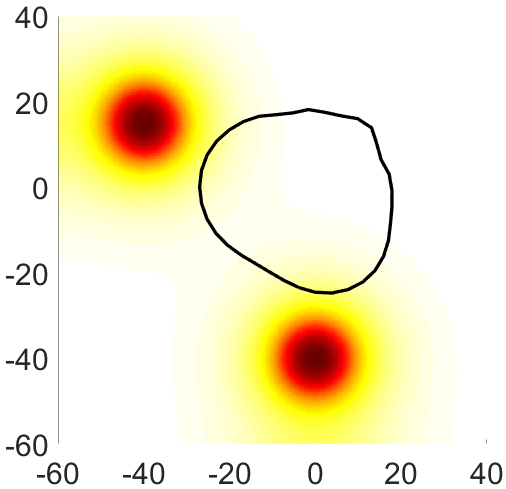}
		& \includegraphics[height=10em]{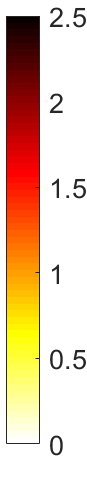}
		\\
		%		& \includegraphics[height=13.2em]{overlay_chem_new}\\
		\footnotesize{(a) Initial time}
		& \footnotesize{(b) Final time}
	\end{tabular}		
	\caption{Initial and final time conformation of the Experiment \ref{exp:sensit} (sensitivity analysis). Showing here the cell (solid line) and the two pipettes as circular sources of chemical. (a): The initial condition of the cell and the chemical. (b): In the final time the cell has responded to the chemical gradient by breaking its symmetry and protruding towards both sources of the chemical. The colorbar in the right refers to the density of the chemical and is common to both figures.}\label{fig:sensit}
\end{figure}

\begin{experiment}[Sensitivity -- Chemotaxis]\label{exp:sensit}
	An initially rotational symmetric cell resides over an adhesively uniform substrate 
	\begin{equation}
		\mu^A(\vec x) = 0.75\,.
	\end{equation}
	Two sources of \change{}{the} chemical are found in the vicinity of the cell as seen Figure \ref{fig:sensit} (a). They represent two \textit{pipettes} of the same circular shape that inject \change{}{the} chemical in the environment with the same rate, cf. Tables \ref{tbl:parameters} and  \ref{tbl:env.parameters} for the relevant parameters.
\end{experiment}
	In Figure \ref{fig:sensit} we reproduce the time evolution of the Experiment \ref{exp:sensit}. As the chemical diffuses in the environment, it decays with time (chemical degradation), and is degraded upon attachment with the cell. The cell responds to the gradient of the chemical by adjusting its polymerization rate, breaking \change{of}{} its symmetry, and moving towards the direction of the pipettes. 
	
The parameters we consider in the sensitivity analysis of the FBLM-FEM are
\begin{equation}\label{eq:sens.set}
	\mathcal P^{\text{sens}}=\Big\{\mu^B, \mu^A, \mu^T, \mu^S, \phi_0, \mu^{IP}, v_{\min}, v_{\max}, \lambda_\text{inext}, \mu^P, \lambda_\text{tether}, A_0\Big\}
\end{equation}
that we index by $i= 1\dots 12$ and set the \textit{reference parameter set} to be  
\begin{equation}\label{eq:ref.set}
	\mathcal P^\text{ref}=\Big\{ p_i^\text{ref},\ i=1\dots 12\Big\}
\end{equation}
with values given in Table \ref{tbl:parameters}. For this parameter set, the final time conformation of the cell is denoted by $\mathcal C^\text{ref}$ and is depicted in Figure \ref{fig:sensit} (b).

We perturb one after the other the reference parameters $p_i^\text{ref}\in \mathcal P^\text{ref}$ to new values $p_i^\text{per}$, while maintaining the rest to their reference values. For each perturbation, the new parameter set $\mathcal P^\text{per}_i$ differs from the reference set $\mathcal P^\text{ref}$ only at the parameter $i$. 

For each perturbation $\mathcal P^\text{per}_i$ of the parameter set, we compute the final time conformation of the cell $\mathcal C^\text{per}_i$ and compare it with the reference $\mathcal C^\text{ref}$ as:
\begin{equation}\label{eq:sensit}
	\mathcal S^{|\cdot|_X}_{i}=\frac{|\mathcal C^\text{per}_i|_X - |\mathcal C^\text{ref}|_x }{p_i^\text{per}-p_i^\text{ref}}.
\end{equation}
where the ``norm'' $|\cdot|_X$, $X=A,B,C,D$ is one of the ``norms'' introduced in Section \ref{sec:converge}. The perturbations of the parameters are small, hence the divided differences in \eqref{eq:sensit} can also be viewed as approximations to the corresponding derivatives, around the reference state $\mathcal P^\text{ref}$.

\begin{figure}[t]
	\centering
	\footnotesize
	\begin{tabular}{>{\centering}m{0.9\linewidth}}
		\includegraphics[width=1\linewidth]{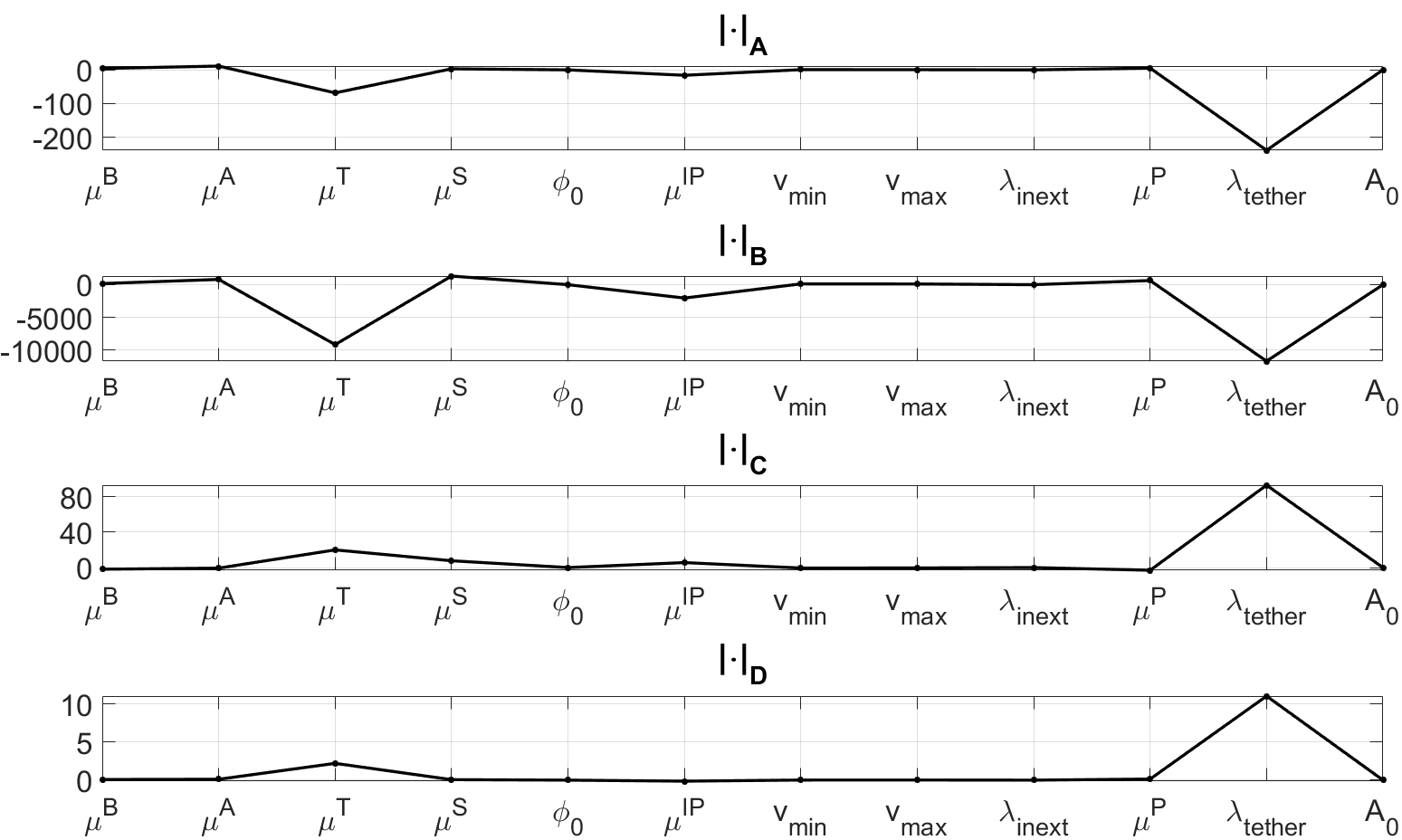}\\
		(a)	Sensitivity results for all the parameters in $\mathcal P^\text{sens}$ \eqref{eq:sens.set}.\\
		\includegraphics[width=1\linewidth]{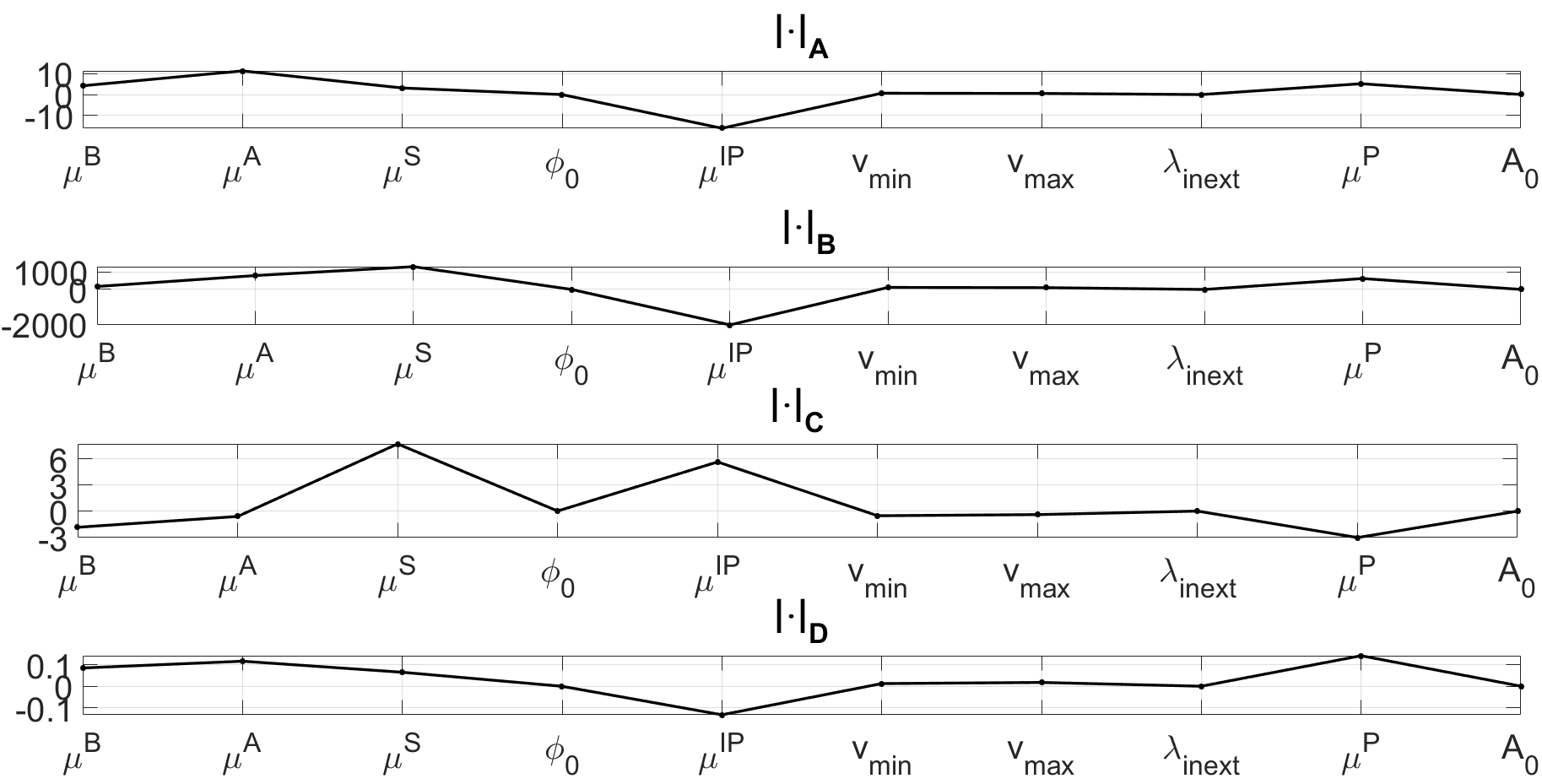}\\
		(b) Subtracting the most influential parameters $\mu^T$ and $\lambda_\text{tether}$ from (a), reveals the importance of the adhesion coefficient $\mu^A$, the inner pulling force $f_\text{inn}$, the pressure $\mu^P$.
	\end{tabular}
	\caption{Graphical representation of the sensitivity analysis Experiment \ref{exp:sensit}. The effect of all the parameters \eqref{eq:sens.set} is shown in the panel (a). Clearly, the membrane tethering $\lambda_\text{texther}$ and the twisting coefficient $\mu^T$ are the more influential parameters. In (b) we exclude these two parameters and can hence identify the stretching $\mu^S$, inner-pulling $\mu^\text{IP}$, and the pressure $\mu^P$ as the (next) more influential parameters, cf.  Table \ref{tbl:sensit}. The ``norms'' $|\cdot|_A$, $|\cdot|_B$, $|\cdot|_C$, and $|\cdot|_D$ represent: \change{I}{i}nvasiveness, \change{S}{s}ize, \change{P}{p}erimeter, and \change{E}{e}longation \change{}{of the cell as defined in \eqref{eq:norm.invad}-\eqref{eq:norm.elong}}.}\label{fig:sensit_graph}
\end{figure}

In essence, $\mathcal S_i^{|\cdot|_X}$ represents the rate at which the cell changes, in the sense of the ``norm'' $|\cdot|_X$, with respect to the parameter $i$. Accounting for all the parameters of $\mathcal P^\text{sens}$, the \textit{local sensitivity} follows. We refer to the Tables and \ref{tbl:parameters} for the full list of the reference parameters and to Table \ref{tbl:sensit} and Figure \ref{fig:sensit_graph} for a concise description of the sensitivity analysis results. 

We note that these results are not global in the sense that they are influenced by the experiment under investigation, the initial state of the cell, the environment, the reference parameters, and more. %This can be seen in the following example: with the same preferred inner area $A_0$ the effect of the actin-myosin pulling force is expected to be higher if the initial cell is larger rather than when it is smaller. 

Nevertheless, we present here some characteristic remarks that will assist in further investigations:
\begin{itemize}
	\item[--] The membrane tethering parameter $\lambda_\text{tether}$, is the most influential of the parameters \eqref{eq:sens.set} in all the ``norms''. The fact in particular that it has a negative effect in the \textit{invasiveness} $|\cdot|_A$ ``norm'' and a positive in the \textit{perimeter} $|\cdot|_C$ and \textit{elongation} $|\cdot|_D$ ``norms'', implies that its primer effect is retractive not on the protruding but rather on the side parts of the cell.
	
	\item[--] The twisting parameter $\mu^T$ is equally important. This is in contrast to the lesser influence of the $\mu^S$ parameter (the other biological component of the crosslink protein). Moreover, the negative influence in \textit{area} $|\cdot|_B$ and positive in \textit{perimeter} $|\cdot|_C$ ``norms'' is understood by the fact that the decrease of $\mu^T$ leads to more linear and radial filaments, and to rotational symmetric and circular cells; hence to (relative) increase of the area and decrease of the perimeter. The decrease of the preferred angle $\phi_0$ has also the same effect.

	\item[--] All the parameters, except for $A_0$, have opposite effects in the \textit{area} $|\cdot|_B$ and \textit{perimeter} $|\cdot|_C$ ``norms''. As in the case of $\mu^T$, this is understood by the fact that each term either leads towards to or away from a more circular conformation of the cell, which in turn maximizes the area and minimizes the perimeter. In contrast, \change{the} increase of $A_0$ leads to a larger cell by increasing the inner area of the cell (behind the lamellipodium) and which leads to increase of the total area of the cell and its perimeter.
	
	\item[--] Increasing the polymerization rates $v_{\min}$ and $v_{\max}$ has a positive effect in all the ``norms'' except for the \textit{perimeter} $|\cdot|_C$. This is so since the polymerization of filaments opposes the retracting effects of the other components of the model which in turn are responsible for the deformation of the initial rotational symmetric cell; in short: increasing the polymerization rates, leads to more circular cells. 
	
	\item[--] The myosin-actin inner pulling parameter $\mu^\text{IP}$ is also very influential in all the ``norms''. Note also that $\mu^\text{IP}$ and $\phi_0$ are the only parameters with negative impact in the \textit{elongation} ``norm'' $|\cdot|_D$. That is, the higher the $\mu^\text{IP}$ or the $\phi_0$ are, the less elongated the cell becomes, see also Figure \ref{fig:softstiff}.
	
	\item[--] We note the effect of $\mu^A$ and $\mu^S$ is almost identical in the \textit{invasiveness} and \textit{elongation} ``norms'' $|\cdot|_A$, $|\cdot|_D$. On the other hand, in the \textit{size} and \textit{elongation} ``norms'' $|\cdot|_B$, $|\cdot|_C$, their effects are of similar magnitude but opposite sign. This is understood as follows: \change{the} increase of the adhesion parameter $\mu^A$ leads to an increase of the size of the cell (see also \cite{Brunk2016}) whereas \change{the} increase of the stretching parameter $\mu^S$ leads to its decrease.
\end{itemize}

\begin{table}[t]
	\footnotesize
	\centering
	\begin{tabular}{c|p{7em}|p{7em}|p{7em}|p{7em}}
		&\centering$|\cdot|_A$&\centering$|\cdot|_B$&\centering$|\cdot|_C$&~\hfill$|\cdot|_D$\hfill~\\ 
		variable&\centering{Invasiveness}&\centering{Area}&\centering{Perimeter}&~\hfill{Elongation}\hfill~\\
		\hline
		&&&&\\[-0.7em]
		$\mu^B$&$+4.1903$&$+5.4005\times 10^2$&$-1.8422$&$+8.4974\times 10^{-2}$\\
		$\mu^A$&$+1.1291\times 10^{1}$&$+1.2492\times 10^{3}$&$-6.0469\times 10^{-1}$&$+1.1648\times 10^{-1}$\\
		$\mu^T$&$+6.8379\times 10^{1}$&$-6.3180\times 10^{3}$&$+1.9945\times 10^{1}$&$+2.2043$\\
		$\mu^S$&$+3.1246$&$-1.5439\times 10^{3}$&$+7.6924$&$+6.5057\times 10^{-2}$\\
		$\phi_0$&$-1.6577\times 10^{-2}$&$-9.6335$&$+1.6478\times 10^{-2}$&$-2.8298\times 10^{-4}$\\
		$\mu^\text{IP}$&$-1.6111\times 10^{1}$&$-2.7006\times 10^{3}$&$+5.6310$&$-1.3288\times 10^{-1}$\\
		$v_\text{min}$&$+6.3210\times 10^{-1}$&$+1.5715\times 10^{2}$&$-5.2289\times 10^{-1}$&$+1.2388\times 10^{-2}$\\
		$v_\text{max}$&$+5.2297\times 10^{-1}$&$+1.6010\times 10^{2}$&$-3.9309\times 10^{-1}$&$+1.7660\times 10^{-2}$\\
		$\lambda_\text{inext}$&$-1.9399\times 10^{-2}$&$-4.5358$&$+8.7613\times 10^{-3}$&$+9.6935\times 10^{-6}$\\
		$\mu^P $&$+5.1689$&$+1.5652\times 10^{3}$&$-3.0199$&$+1.4201\times 10^{-1}$\\
		$\lambda_\text{tether}$&$-2.3896\times 10^{2}$&$-9.5477\times 10^{4}$&$+9.2927\times 10^{1}$&$+1.1029\times 10^{1}$\\
		$A_0$&$+1.7537\times 10^{-2}$&$+2.1346$&$+1.5987\times 10^{-4}$&$+3.1229\times 10^{-4}$\\		
	%	$\kappa_\text{ref}$ & reference leading edge curvature for polymerization speed reduction & \\
		%\noalign{\smallskip}
%		\hline
	\end{tabular}
	\caption{Sensitivity of the FBLM-FEM with respect to the parameters \eqref{eq:sens.set} in Experiment \ref{exp:sensit}. Larger absolute values imply stronger influence of the corresponding parameters (rows) to the different ``norms''(columns). See Section \ref{sec:sensit} for the computation and discussion of these results.}
	\label{tbl:sensit}
\end{table}    
\section{Discussion}

The aim of this work was twofold: to investigate two fundamental numerical properties (stability and convergence) of the FEM solving the FBLM, and to embed the FBLM in a complex and adaptive extracellular environment and study its sensitivity to several of its controlling parameters. \change{In some more detail:}{}

We have showed in Section \ref{sec:stbility} that the FEM exhibits the expected timestep stability behaviour \eqref{eq:CFL} in both chemotaxis and haptotaxis experiments. We have verified this assertion in a particular chemotaxis experiment where we have identified the stability constant $C$ and have proposed an automated way of computing proper timesteps for the method. The technique proposed is inspired by the well known ATC methods used in many cases of scientific computing, and is based on the smoothness of the filament polymerization rate and the filament density functions. 

We have proceeded further and identified with a series of haptotaxis experiments the dependence between the timestep stability and the gradient of the ECM. We have seen that this relation \eqref{eq:C.hapt} is linear and we expect it to hold globally, although with different coefficients.

We have exhibited in Section \ref{sec:converge} the convergence of the FEM as the discretization grid is refined with respect to $\Delta \alpha$ and/or $\Delta s$. As is commonly done in complex models/methods, we have considered characteristic/representative experimental cases. Since the exact solutions are unknown we deduce the convergence by comparison against the numerical solutions of the finest discretization grid. The comparisons themselves are conducted in terms of particular ``norms'' that we define in  \eqref{eq:norm.invad}--\eqref{eq:norm.elong}.

In Section \ref{sec:env} we have dealt with the embedding of the FBLM-FEM in a complex and adaptive extracellular environment. The model for the environment that we propose in \eqref{eq:env} is an reaction-diffusion system of the densities of the involved quantities. The proposed model is relatively simple and is used here mostly to exhibit the coupling between the FBLM and the environment. 

We close this work in Section \ref{sec:sensit} with a first study of the sensitivity of the FBLM-FEM on several of its controlling parameters. This sensitivity analysis is performed around predefined parameter values, most of which are biologically relevant and that have been previously proposed in the literature. We identify the most significant parameters and get an insight on the magnitude of the effect of the different model components.

Although the current work is based on particular experimental cases and is not escorted by rigorous numerical (or other type of) analysis, the benefit is twofold: 

On the one hand, the combination FBLM-FEM can be further used to model and simulate biologically relevant experimental situations. We know now that the refinement of the discretization grid, augmented by the automated adaptation of the timestep, provides with a stable and converging method that will reveal the inherent dynamics of the model. 

On the other hand, having verified that the behaviour of the FBLM-FEM is the expected one in terms of convergence and stability, the rigorous numerical analysis is well warranted. It is expected that the remarks of this work will serve also as a guide in this effort.

Moreover, the coupling of the FBLM with the extracellular environment  that we propose here will serve as springboard for the further modelling and simulation of more biologically relevant settings and \textit{in-vitro} experiments. It can easily be extended to include e.g. the description of the ECM as a fibrous component of the environment, more than one chemical ingredients (attractants or repellents), as well as more than one cells interacting with each other. 

The sensitivity analysis results have served here for the deeper understanding of the effect of different terms on the model. The insight we have gained will serve in the further refinement of the FBLM-FEM and the parameter estimation procedures, when reproducing and simulating realistic experimental scenarios.

%-----------------------------------------------------------------------------
%----------- for better visual results, we put here all the parameter tables
%-----------------------------------------------------------------------------

\begin{table}[t]
	\footnotesize
	\centering
	\begin{tabular}{ c p{0.29\linewidth} |p{0.24\linewidth}| p{0.27\linewidth}}

		symb.&\hfill~description\hfill~&\hfill~value\hfill~&\hfill~comment\hfill~\\ 
		\hline
		& & & \\
		$\mu^B$&bending elasticity& $\rm 0.07\,pN\,\upmu m^2$ &\cite{Gittes1993}  \\		
		$\mu^A$&adhesion& $\rm 0.4101\,pN\, min\, \upmu m^{-2}$& \cite{Li2003,Oberhauser2002} \& \cite{Oelz2008,Oelz2010a,Schmeiser2010}\\
		$\mu^T$&cross-link twisting& $\rm 7.1\times 10^{-3}\, \upmu m$ &\\
		$\mu^S$&cross-link stretching& $\rm 7.1 \times 10^{-3}\, pN\, min\, \upmu m^{-1} $& \\
		$\phi_0$&crosslinker equil. angle& $70^o$ & \cite{Schmeiser2010}\\
%		$\kappa_\text{br}$ & branching rate & $\rm 10\, min^{-1} $ & order of magnitude from \cite{Grimm2003}, chosen to fit $2 \overline \rho_\text{ref}=\rm 90\, \upmu m^{-1}$ \cite{Small2002}\\
%		$\kappa_\text{cap}$ & capping rate & $\rm 5\, min^{-1} $ & order of magnitude from \cite{Grimm2003}, chosen to fit $2 \overline \rho_\text{ref}=\rm 90\, \upmu m^{-1}$ \cite{Small2002}\\
%		$c_\text{rec}$ & Arp$2/3$ recruitment  & $\rm 900\, \upmu m^{-1}\, min^{-1} $ & chosen to fit $2 \overline \rho_\text{ref}=\rm 90\, \upmu m^{-1}$ \cite{Small2002}\\
%		$\kappa_\text{sev}$ & severing rate & $\rm 0.38\, min^{-1}\, \upmu m^{-1}$ & chosen to give lamellipodium widths similar as described in \cite{Small2002} \\
		$\mu^\text{IP}$&actin-myosin strength&$\rm 0.1\, pN\, \upmu m^{-2}$& \\
		$v_\text{min}$&minimal polymerization&$\rm 1.5\,\upmu m\, min^{-1}$&in biological range\\
		$v_\text{max}$&maximal polymerization&$\rm 8\,\upmu m\, min^{-1}$&in biological range\\
		$\mu^P $ &pressure constant & $\rm 0.05\, pN\, \upmu m $& \\
%		$\kappa_\text{ref}$ & reference leading edge curvature& $\rm (5\,\upmu m)^{-1}$ &\\
		$A_0$ & equilibrium inner area & $\rm 450\, \upmu m^2$&\cite{Verkhovsky1999,Small1978}\\
		\hline
		$\lambda_\text{inext}$&inextensibility&$20$&\\
		$\lambda_\text{tether}$&membrane tethering&$1\times 10^{-3}$&\\
		%\noalign{\smallskip}
	\end{tabular}
	\caption{Basic set of parameter values used in the numerical simulations of the FBLM. Variants of these parameters are discussed in each experiment separately. These parameters (except for the $\lambda_\text{inext}$ and $\lambda_\text{tether}$) have been adopted from \cite{MOSS-numeric}.}
	\label{tbl:parameters}      
\end{table}
 
\begin{table}[t]
	\footnotesize
	\centering
	\begin{tabular}{c p{12em} |p{8em}|c|c}
		symb.&\hfill~description\hfill~ &\hfill~Experiment \ref{exp:embedding}\hfill~& Experiment \ref{exp:softstiff}  & Experiment \ref{exp:sensit}\\ 
		&&\hfill~and Figure \ref{fig:embedding}\hfill~&and Figure \ref{fig:softstiff}&and Figure \ref{fig:sensit}\\
		\hline
		& & & \\[-0.7em]
		$D_c$  &diffusion of the chemical& $2\times10^3\, {\rm cm^2min^{-1}}$ & $2\times10^3$& $5\times10^3$\\           % diffusion of both chemicals
		$D_m$ &diffusion of the MMPs& $1\, {\rm cm^2min^{-1}}$ & $2\times10^3$ & $5\times10^3$\\
		$\alpha_1$ &production rate of chemical& $10^2\, {\rm mol\,min^{-1}}$ & $10^2$ & $4\times10^2$\\      % production of extracell. chemical by source
		$\beta$ & production of MMPs& $10^{-1}\, {\rm mol\,min^{-1}}$ & $0$ & $0$\\      % production of MMPs by the cell 
		$\gamma_1$ &decay of the chemical& $10\, {\rm mol\,min^{-1}}$ & $10$& $7\times 10^1$\\       % decay of extracell. chemical
		$\gamma_2$ &decay of the MMPs& $10\, {\rm mol\,min^{-1}}$& $0$& $0$\\       % degradation of MMPs
		$\delta_1$ &degr. chemical by the cell& $0$ & $0$ & $0$\\      % degradation of the extracell. chemical by the cells
		$\delta_2$ &degr. of the ECM by the MMPs&$2\,{\rm cm^2mol^{-1}min^{-1}}$& $5\times10^{-1}$& $0$    % degradation of the ECM
	\end{tabular}
	
	\caption{Parameter sets used for the simulation of the environment \eqref{eq:env} in the Experiments \ref{exp:embedding}, \ref{exp:softstiff}, \ref{exp:sensit}, see also Figures \ref{fig:embedding}, \ref{fig:softstiff}, \ref{fig:sensit}. There is no biological justification of these values.}
	\label{tbl:env.parameters}      
\end{table}
 
%----------- Acknowledgement
\section*{Acknowledgement}
The authors would like to thank Christian Schmeiser, Anna Marciniak-Czochra, and Mark Chaplain for the fruitful discussions and suggestions during the preparation of this manuscript.

%-----------------------------------------------------------------------------
%-----------------------------------------------------------------------------
%\appendix
%\section{Parameters for the experiments}
%\subsection{Parameters for ECM degradation}~
%
%\begin{table}[h]
%	\footnotesize
%	\centering
%	\begin{tabular}{c p{12em} ccc}
%		\hline
%		var. & meaning &Experiment \ref{exp:embedding} & Experiment \ref{exp:softstiff}  & Experiment \ref{exp:sensit}\\ 
%		&&and Figure \ref{fig:embedding}&and Figure \ref{fig:softstiff}&and Figure \ref{fig:sensit}\\
%		\hline
%		& & & \\[-0.7em]
%		$D_c$  &diffusion of the chemical& $2000$ & $2000$& $5000$\\           % diffusion of both chemicals
%		$D_m$ &diffusion of the MMPs& $1$ & $2000$ & $5000$\\
%		$\alpha_1$ &production rate of chemical& $100$ & $100$ & $400$\\      % production of extracell. chemical by source
%		$\beta$ & production of MMPs& $0.1$ & $0$ & $0$\\      % production of MMPs by the cell 
%		$\gamma_1$ &decay of the chemical& $0.1$ & $10$& $70$\\       % decay of extracell. chemical
%		$\gamma_2$ &decay of the MMPs& $10$& $0$& $0$\\       % degradation of MMPs
%		$\delta_1$ &degr. chemical by the cell& $0$ & $0$ & $0$\\      % degradation of the extracell. chemical by the cells
%		$\delta_2$ &degr. of the ECM by the MMPs& 2& $0.5$& $0$    % degradation of the ECM
%	\end{tabular}
%	\caption{~}
%\end{table}

%----------------------------------------------------------------------------
	\bibliographystyle{unsrtnat}%{plain}%{alpha} %{agsm}%{plain}%{amsalpha}
%\begin{multicols}{2}
{	\footnotesize
	\bibliography{FBLM-FEM-numerics}
}
%\end{multicols}

%------------------------------------------------------------------------------------------
%------------------------------------------------------------------------------------------
%------------------------------------------------------------------------------------------
\appendix
\section{The FEM for the FBLM} \label{sec:FEM}
We numerically solve the FBLM \eqref{eq:strong} with a problem specific FEM that was first presented in  \cite{MOSS-numeric}. Here we present some of its components.

The maximal filament length varies around the lamellipodium, and in effect the computational domain 
\[
B(t) = \left\{(\alpha,s):\, 0\le\a<2\pi\,,\, -L(\alpha,t)\le s < 0\right\}
\]
is non-rectangular.  For consistency and stability reasons we recover the orthogonality of the domain $B(t)$, using the coordinate transformation 
\[
(\alpha,s,t) \rightarrow \(\alpha, L(\alpha,t)s, t\) \,,
\]
and replace it by
\begin{equation}\label{eq:rescaled.B}
B_0 := [0,2\pi)\times[-1,0)\ni (\alpha,s)\,.
\end{equation}
%\begin{figure}[t]
%	\centering 	
%	\begin{tabular}{ccc}
%		{\includegraphics[height=12em]{input/DisretizedCell.pdf}}
%		&&{\includegraphics[height=10em]{input/IndexedCellPart.pdf}}\\
%		\footnotesize{(a)} && \footnotesize{(b)}
%	\end{tabular}
%	\caption{A rotational symmetric lamellipodium along with the projection of both families of discretization filaments of the domain $B_0$. In (a) are shown the ``discretization filaments'' of the complete lamellipodium, and in (b) a lamellipodium fragment detail with the enumeration of the filaments and of the discretization cells.}\label{fig:discr}
%\end{figure}

Accordingly, the weak formulation of (\ref{eq:strong}), recasts into
\begin{align}
0 	=&\int_{B_0}\eta \left( \mu^B \partial^2_s\vec{F} \cdot \partial_s^2\vec{G} + L^4\mu^A \widetilde{D_t} \vec{F} \cdot \vec{G} + L^2\lambda_\text{inext} \partial_s\vec{F} \cdot \partial_s\vec{G} \right) d(\alpha,s) \nonumber\\
&+\int_{B_0}  \eta\eta^*\left( L^4\widehat{\mu^S} \(\widetilde{D_t}\vec{F}  - \widetilde{D_t^*}\vec{F}^* \)\cdot \vec{G} \mp L^2\widehat{\mu^T}(\phi-\phi_0)\partial_s\vec{F}^{\perp} \cdot \partial_s \vec{G} 
\right) \ d(\alpha,s) \nonumber\\
&- \int_{B_0} p(\varrho) \left( L^3\partial_\a \vec{F}^{\perp}\cdot \partial_s \vec{G} - \frac{1}{L}\partial_s \vec{F}^{\perp}\cdot \partial_\a (L^4\vec{G})\right) d(\alpha,s) \nonumber\\
&+	\int_0^{2\pi} \eta \left( L^2 f_\text{tan}\partial_s\vec{F} + L^3 f_{\text{inn}}\vec{V} \right)\cdot \vec{G} \Bigm|_{s=-1}d\a \mp	\int_0^{2\pi} L^3\lambda_\text{tether}\nu\cdot \vec{G} \Bigm|_{s=0} d\a \,,\label{eq:FEM}
\end{align}
with $\vec{F}, \vec{G}\in H^1_\a\((0,2\pi);\,H^2_s(-1,0)\)$. In a similar manner the modified material derivative and in-extensibility conditions read 
\[
\widetilde{D_t} = \partial_t - \(\frac{v}{L} + \frac{s \partial_t L}{L}\)\partial_s
\]
and
\[
\left|\partial_s \mathbf F(\alpha,s,t)\right| = L(\alpha,t).
\]

We decompose $B_0$ into disjoined rectangular \textit{computational cells} as follows:
\begin{equation}\label{eq:cell}
B_0 = \bigcup_{i=1}^{N_a} \bigcup_{j=1}^{N_s -1} C_{i,j} \,,\quad
\mbox{where}\quad  C_{i,j}=[\alpha_i,\alpha_{i+1})\times[s_j,s_{j+1}) \,,
\end{equation}
for $\alpha_i = (i-1)\Delta \alpha$, $\Delta \alpha = \frac{2\pi}{N_\a}$, $i = 1,\ldots,N_\a+1$, and $s_j = -1 + (j-1)\Delta s$, $\Delta s = \frac{1}{N_s - 1}$, $j = 1,\ldots,N_s$. The resolution of the grid along the $\a$ and $s$ directions is denoted by $\anodes$, $\snodes$. The $\a$-periodicity assumption suggests that $\alpha_{N_\a+1} = 2\pi$ is identified with $\alpha_1 = 0$.%, see also Figure \ref{fig:discr}.

We follow \cite{MOSS-numeric} and set the conforming FE space
\begin{align}
\mathcal V := \Bigl\{ \vec{F}\in C_\a&\([0,2\pi];\, C^1_s([-1,0])\)^2 \text{ such that } \vec{F}\bigm|_{C_{i,j}}(\cdot,s) \in \mathbb{P}^1_\a\,,\, \nonumber \\
&\vec{F}\bigm|_{C_{i,j}}(\a,\cdot) \in \mathbb{P}^3_s\quad\mbox{for } i=1,\ldots,N_\a\,;\,  j = 1,\ldots,N_s-1\Bigr\}  \,, \label{eq:FE:V}
\end{align}
of continuous functions that are continuously differentiable with respect to $s$, and such that on each computational cell they coincide with a first order polynomial in $\a$, and a third order polynomial in $s$.

%\begin{figure}[t]
%	\centering
%	\begin{tabular}{cccc}
%		{\includegraphics[width=0.21\textwidth]{input/H1.pdf}}&
%		{\includegraphics[width=0.21\textwidth]{input/H2.pdf}}&
%		{\includegraphics[width=0.21\textwidth]{input/H3.pdf}}&
%		{\includegraphics[width=0.21\textwidth]{input/H4.pdf}}\\
%		\footnotesize{(a) $H_1^C$} & \footnotesize{(b) $H_1^C$} & \footnotesize{(c) $H_1^C$} & \footnotesize{(d) $H_1^C$}
%	\end{tabular}
%	\caption{Graphical representation of four of the Lagrange-Hermite shape functions (\ref{eq:H}) over the generic discretization cell $C$. The $H_1^C$, $H_3^C$, $H_5^C$, and $H_7^C$ attain the value 1 and the derivative 0 at the corners of $C$,  whereas the rest attain the value 1 and derivative 0 at the corners of $C$.}\label{fig:shape}
%\end{figure}

In particular, we consider for $i = 1,\ldots,N_\a+1$, $i = j,\ldots,N_s$, and  $(\a,s)\in C_{i,j}$,  that
\begin{equation}\label{eq:H}
\left\{ \begin{array}{lcl}
H_1^{i,j}(\a,s)=L_1^{i,j}(\a) G_1^{i,j}(s),&~& 
H_5^{i,j}(\a,s)=L_2^{i,j}(\a) G_1^{i,j}(s)\\
H_2^{i,j}(\a,s)=L_1^{i,j}(\a) G_2^{i,j}(s),&& 
H_6^{i,j}(\a,s)=L_2^{i,j}(\a) G_2^{i,j}(s)\\
H_3^{i,j}(\a,s)=L_1^{i,j}(\a) G_3^{i,j}(s),&& 
H_7^{i,j}(\a,s)=L_2^{i,j}(\a) G_3^{i,j}(s)\\
H_4^{i,j}(\a,s)=L_1^{i,j}(\a) G_4^{i,j}(s),&& 
H_8^{i,j}(\a,s)=L_2^{i,j}(\a) G_4^{i,j}(s) 
\end{array}\right.
\end{equation}
with 
\begin{equation}\label{eq:LG}
\left\{\begin{array}{lcl}
L_1^{i,j}(\a) =\frac{\a_{i+1}-\a}{\Delta \a}, 
&&\quad G_1^{i,j}(s)=1-\frac{3(s-s_j)^2}{\Delta s^2}+\frac{2(s-s_j)^3}{\Delta s^3} \\
L_2^{i,j}(\a)=1-L_1^{i,j}(\a),
&&\quad G_2^{i,j}(s)=s-s_j-\frac{2(s-s_j)^2}{\Delta s}+\frac{(s-s_j)^3}{\Delta s^2}\\
&&\quad G_3^{i,j}(s)=1-G_1^{i,j}(s)\\
&&\quad G_4^{i,j}(s)=-G_2^{i,j}(s_j+s_{j+1}-s)
\end{array}\right.
\end{equation}
and that $H_k^{i,j}(\a,s)=0$, $k=1,\ldots,8$, whenever $(\a,s)\not\in C_{i,j}$.% (see also Figure \ref{fig:shape}). 
The basis functions are then defined as:
\begin{equation}\label{basis}
\left\{\begin{array}{r}
\Phi_{i,j} := H_7^{i-1,j-1}+ H_5^{i-1,j} + H_3^{i,j-1} + H_1^{i,j} \\
\Psi_{i,j} :=H_8^{i-1,j-1}+ H_6^{i-1,j} + H_4^{i,j-1} + H_2^{i,j} 
\end{array}\right.
\end{equation}
for $i=1,\ldots,N_\a,\, j=1,\ldots,N_s$, and the element $\vec{F}\in\mathcal V$ can be represented in terms of the point values $\vec{F}_{i,j}$ and the $s$-derivatives $\partial_s \vec{F}_{i,j}$ at the discretization nodes, as:
\begin{equation}\label{eq:I2D}
\vec{F}(\a,s)=\sum_{i=1}^{N_\a} \sum_{j=1}^{N_s} \big( \vec{F}_{i,j} \Phi_{i,j}(\a,s) + \partial_s\vec{F}_{i,j} \Psi_{i,j}(\a,s) \big) \,.
\end{equation}

The FE formulation of the lamellipodium problem on the time interval $[0,T]$ is to find  $\vec{F}\in C^1\big([0,T];\,\mathcal V\big)$, such that (\ref{eq:FEM}) holds for all $\vec{G}\in C\big([0,T];\,\mathcal V\big)$.

%-----------------------------------------------------------------------------------------------------
%-----------------------------------------------------------------------------------------------------

\section{The FV method the environment}\label{sec:FV}
We solve the \eqref{eq:env} using a FV method that was previously developed in \cite{Sfakianakis.2016, Sfakianakis.2016b} where we refer for details. Here we provide some information.

We consider the \textit{advection-reaction-diffusion} (ARD) system 
\begin{equation}\label{eq:app.gen.PDE}
\vec w_t = A(\vec w) + R(\vec w) + D(\vec w),
\end{equation}
where $\vec w$ represents the solution vector, and $A$, $R$, and $D$ the \textit{advection}, \textit{reaction}, and \textit{diffusion} operators respectively. 

We denote by $\vec w_h(t)$ the corresponding (semi-)discrete numerical approximation, indexed by the maximal diameter of the spatial grid $h$,  that satisfies the system of ODEs
\begin{equation}\label{eq:app.gen.scheme}
\partial_t \vec w_h = \mathcal A(\vec w_h) + \mathcal R(\vec w_h) + \mathcal D(\vec w_h),
\end{equation}
where the numerical operators $\mathcal A$, $\mathcal R$, and $\mathcal D$ are \textit{discrete approximations} of the operators $A$, $R$, and $D$ in \eqref{eq:app.gen.PDE} respectively. 

We \textit{split} \eqref{eq:app.gen.scheme} in an \textit{explicit} and an \textit{implicit} part as 
\begin{equation}\label{eq:app.gen.IMEX}
\partial_t \vec w_h = \mathcal I(\vec w_h) + \mathcal E(\vec w_h).
\end{equation}
The details of the splitting depend on the particular problem in hand but in a typical case, the advection terms $\mathcal A$ are explicit in time, the diffusion terms $\mathcal D$ implicit, and the reaction terms $\mathcal R$ partly explicit and partly implicit, according to the reaction rates.

More precisely, we employ a diagonally implicit RK method for the implicit part, and an explicit RK for the explicit part
%\begin{eqnarray}
%\begin{equation*}%\label{F:eq:IMEX3}
%\begin{array}{l}
%	\vec w_h^{n+1}=\vec w_h^n +\tau_n\Big( \sum\limits_{j=1}^{i-1}b_{j}^E ({\rm -A + R_\text{expl}})(t_n + c_j^E \tau, \vec W_i)
%+\sum\limits_{j=1}^{i}b_{j}^I({\rm D + R_\text{impl}})(\vec W_i) \Big),\\
%\vec W_i= \vec w_h^n +\tau_n \Big( \sum\limits_{j=1}^{i-1}a_{ij}^E ({\rm -A + R_\text{expl}})(t_n + c_j^E \tau, \vec W_i)
%+\sum\limits_{j=1}^{i}a_{ij}^I(\rm D + R_\text{impl})(\vec W_i) \Big) . 
%\end{array}
%\end{equation*}		
%\end{eqnarray}
\begin{equation}\label{eq:app.IMEXRK}
\begin{cases}
\vec W_i^\ast = \vec w_h^n + \tau_n \sum_{j=1}^{i-2}\bar a_{i,j}\vec E_j + \tau_n \bar a_{i,i-1}\vec E_{i-1},&\quad i=1\dots s\\
\vec W_i = \vec W_i^\ast + \tau_n \sum_{j=1}^{i-1} a_{i,j}\vec I_j + \tau_n a_{i,i}\vec I_i,&\quad i=1\dots s\\
\vec w_h^{n+1} = \vec w_h^n + \tau_n \sum _{i=1}^s\bar b_i \vec E_i + \tau_n \sum_{i=1}^sb_i\vec I_i
\end{cases},
\end{equation}
where $s=4$ are the stages of the IMEX method, $\vec E_i=\mathcal E(\vec W_i)$, $I_i=\mathcal I(\vec W_i)$, $i=1\dots s$, $\{\bar b,\, \bar A\}$, $\{b,\, A\}$ are respectively the coefficients for the explicit and the implicit part of the scheme, given in the Butcher Tableau in Table \ref{F:tbl:IMEX},  \cite{christopher2001additive}. The linear systems in \eqref{eq:app.IMEXRK} are solved using the \textit{iterative biconjugate gradient stabilized Krylov subspace} method \cite{Krylov.1931, vdVorst.1992}.

\begin{table}[t]
	\centering
	\begin{tabular}{c|cccc}
		$0$&&&&\\[0.5em]
		$\frac{1767732205903}{2027836641118}$&$\frac{1767732205903}{2027836641118}$&&&\\[0.5em]
		$\frac{3}{5}$&$\frac{5535828885825}{10492691773637}$&$\frac{788022342437}{10882634858940}$&&\\[0.5em]
		$1$&$\frac{6485989280629}{16251701735622}$&$-\frac{4246266847089}{9704473918619}$&$\frac{10755448449292}{10357097424841}$&\\[0.5em]
		\hline\\[-0.5em]
		&$\frac{1471266399579}{7840856788654}$ & $-\frac{4482444167858}{7529755066697}$ & $\frac{11266239266428}{11593286722821}$ & $\frac{1767732205903}{4055673282236}$ 
	\end{tabular}
	%	\\[0.5em]
	%	\footnotesize{\textit{(a):}}
	%\caption{Butcher tableau for the explicit part of a third order IMEX scheme, \cite{christopher2001additive}.}
	%\label{F:tbl:IMEX3expl}
	%\end{table}
	\\[0.5em]
	%%\renewcommand{\arraystretch}{1.5}
	%\begin{table}[t]
	\begin{tabular}{c|cccc}
		$0$&0&&&\\[0.5em]
		$\frac{1767732205903}{2027836641118}$&$\frac{1767732205903}{4055673282236}$&$\frac{1767732205903}{4055673282236}$&&\\[0.5em]
		$\frac{3}{5}$&$\frac{2746238789719}{10658868560708}$&$-\frac{640167445237}{6845629431997}$&$\frac{1767732205903}{4055673282236}$&\\[0.5em]
		$1$&$\frac{1471266399579}{7840856788654}$&$-\frac{4482444167858}{7529755066697}$&$\frac{11266239266428}{11593286722821}$&$\frac{1767732205903}{4055673282236}$\\[0.5em]
		\hline\\[-0.5em]
		&$\frac{1471266399579}{7840856788654}$ & $-\frac{4482444167858}{7529755066697}$ & $\frac{11266239266428}{11593286722821}$ & $\frac{1767732205903}{4055673282236}$ 
	\end{tabular}
	\caption{Butcher tableaux for the explicit (upper) and the implicit (lower) parts of the third order IMEX scheme \eqref{eq:app.IMEXRK}, see also \cite{christopher2001additive}.}
	\label{F:tbl:IMEX}
\end{table}

%----------- Supplementary
%\section*{Supplementary material}

%\remm{N@A: should we include supplementary material in the form of animations... e.g. the degradation experiment seen in Fig. 4}
\end{document}